\documentclass{aa}
\usepackage{graphicx}
\def\cs{CS\,31082-001\,}
\def\sned{CS\,22892-052\,}
\def\hd115{HD\,115444\,}
\begin{document}
   \title{The extreme $r$-element rich, iron-poor halo giant CS~31082-001}

   \subtitle{Implications for the $r$-process site(s) and radioactive
              cosmochronology
\thanks{Based on observations of program 165.N-0276(A) obtained with
              the Very Large Telescope of the 
European Southern Observatory at Paranal, Chile.}
}

   \author{V. Hill \inst{1}
     \and
     B. Plez \inst{2}
     \and
     R. Cayrel \inst{3}
     \and
     T. C. Beers \inst{4}
     \and
     B. Nordstr\"om \inst{5,6}
     \and
     J. Andersen \inst{6}
     \and
     M. Spite \inst{1}
     \and
     F. Spite \inst{1}
     \and
     B. Barbuy  \inst{7}
     \and
     P. Bonifacio \inst{8}
     \and
     E. Depagne \inst{1}
     \and
     P. Fran\c{c}ois \inst{3}
     \and
     F. Primas \inst{9}
}

   \offprints{V. Hill}

   \institute{
             Observatoire de Paris-Meudon, GEPI, 2 pl. Jules Janssen,
             F-92195 Meudon Cedex, France\\
         \email{Vanessa.Hill@obspm.fr}
         \and
             GRAAL, Universit\'e de Montpellier II, F-34095 Montpellier
             Cedex 05, France\\
             \email{Bertrand.Plez@graal.univ-montp2.fr}
         \and
             Observatoire de Paris, GEPI, 61 av. de l'Observatoire,
             F-75014 Paris, France\\
             \email{Roger.Cayrel@obspm.fr}
         \and
             Department of Physics \& Astronomy, Michigan State University,
             East Lansing, MI 48824, USA
         \and
         Lund Observatory, Box 43, S-221 00 Lund, Sweden
         \and
         Astronomical Observatory, NBIfAFG, Juliane Maries Vej 30,
         DK-2100 Copenhagen, Denmark
         \and
    IAG, Universidade de S\~ao Paolo, Departmento de Astronomia, CP
         3386, S\~ao Paulo, Brazil
        \and
    Istituto Nazionale per l'Astrofisica - Osservatorio Astronomico di Trieste,
    Via G.B. Tielpolo 11, I-34131
             Trieste, Italy
        \and 
         European Southern Observatory (ESO),
         Karl-Schwarschild-Str. 2, D-85749 Garching b. M\"unchen
}

   \date{Received January 2002; accepted ...}

\abstract{
We present a high-resolution (R=75,000, S/N $\sim500$) spectroscopic analysis
of the bright (V=11.7), extreme halo giant \cs ([Fe/H] = $-$2.9), obtained in
an ESO-VLT {\em Large Programme} dedicated to very metal-poor stars. We find
\cs to be extremely rich in $r$-process elements, comparable in this respect
only to the similarly metal-poor, but carbon-enriched, giant
\sned. As a result of the extreme overabundance of the heaviest $r$-process
elements, and negligible blending from CH and CN molecular lines, a reliable
measurement is obtained of the U~II line at 386 nm, for the first time in a
halo star, along with numerous lines of Th II, as well as lines of 25 other
$r$-process elements.
Abundance estimates for a total of 43 elements (44 counting Hydrogen) are
reported in \cs, almost half of the entire periodic table.\\
The main atmospheric parameters of \cs are as follows: $T_{eff} = 4825~ \pm 50$
K, $\log g= 1.5\pm 0.3$ (cgs), [Fe/H] = $-$2.9 $\pm 0.1$ (in LTE), and
microturbulence 1.8$\pm$ 0.2 km/s. Carbon and nitrogen are not significantly
enhanced relative to iron. As usual in giant stars, Li is depleted by dilution
($\log$(Li/H)=0.85). The $\alpha$-elements show the usual enhancements with
respect to iron, with [O/Fe]$= 0.6\pm 0.2$ (from [O~I] 6300
\AA), [Mg/Fe]$= 0.45\pm 0.16$, [Si/Fe]$= 0.24\pm 0.1$, and [Ca/Fe]$=
0.41\pm 0.08$, while  [Al/Fe] is near $-$0.5. The {\it r}-process elements show
unusual patterns: among the lightest elements (Z$\sim$40), Sr and Zr follow the
Solar {\it r}-element distribution, but Ag is down by 0.8 dex.
All elements with 56 $\leq$ Z $\leq$ 72 follow the Solar {\it r}-element
pattern, reduced by about 1.25 dex. Accordingly, the [$r$/Fe] enhancement is
about +1.7 dex (a factor of 50), very similar to that of \sned. 
Pb, in contrast, seems to be $below$ the shifted
Solar $r$-process distribution, possibly indicating an error in the latter,
while thorium is more enhanced than the lighter nuclides. In \cs, log(Th/Eu) is
$-0.22\pm 0.07$, higher than in the Solar System ($-$0.46) or in \sned
($-$0.66).  If \cs and \sned have similar ages, as expected for two extreme
halo stars, this implies that the production ratios were different by about 0.4
dex for the two objects.  Conversely, if the Th/Eu production ratio were
universal, an age of 15 Gyr for \sned would imply a {\it negative age} for \cs.
Thus, while a universal production ratio for the $r$-process elements seems to
hold in the interval 56 $\leq$ Z $\leq$ 72, it breaks 
down in the actinide region.\\
When available, the U/Th is thus preferable to Th/Eu for radioactive dating, 
for two reasons: $(i)$ because of its faster decay rate and smaller
sensitivity to observational errors, and $(ii)$ because
the inital production ratio of the 
neighboring nuclides $^{238}$U and $^{232}$Th is more robustly predicted than
the $^{151}$Eu/$^{232}$Th ratio.  Our current best estimate for the age of 
\cs is $14.0\pm 2.4$Gyr. However, the computed actinide production ratios 
should be verified by observations of daughter elements such as Pb and Bi in 
the same star, which are independent of the subsequent history of star 
formation and nucelosynthesis in the Galaxy.
}

   \maketitle
%

\section{Introduction} \label{introduction}

The detailed chemical abundances of the most metal-poor halo stars contain
unique information on the earliest epochs of star formation and nucleosynthesis
in our own and other galaxies. Previous studies of these extreme halo stars
(Bessell \& Norris \cite{bessell84}; Norris et al. \cite{norris01}; see other
references in Table 2 of Cayrel \cite{cayrel96}) have revealed abundance
patterns that are very unlike those found in the disk and in halo stars with
metallicities above [Fe/H] $\sim -2.5$ (e.g., Pagel \& Tautvai\u{s}iene
\cite{pagel97}), both as regards their systematic trends and the dramatic
abundance variations exhibited by some elements.  Recently, a few rare stars
have been found to exhibit large $r$-element enhancements, as compared to Solar
ratios, suggesting that their observed abundances are dominated by the influence
of a single, or at most a very few SNe II progenitors (Sneden et al.
\cite{sneden00}), and opening the way to radioactive dating of these stars, and
of the Galaxy, using the ``Th/Eu chronometer'' (Cowan et al.
\cite{cowan99}; Johnson \& Bolte \cite{johnson01}).  

The difficulty in obtaining the required data for such stars challenges
4m-class telescopes to their limits, hence the accuracy of
the results was significantly limited by the resolution and S/N ratio of the
available spectra. Therefore, we have conducted a large program at the ESO {\it
Very Large Telescope} (VLT) and its UVES spectrograph to obtain a systematic,
homogeneous spectroscopic analysis of a large sample of extreme halo stars from
the HK survey of Beers et al. (\cite{beers92}), and ongoing follow-up work. 
This contribution is the first in a series of several analysis papers
to appear from this program. 

\begin{figure*}
   \centering
   {\includegraphics[angle=-90,width=17cm]{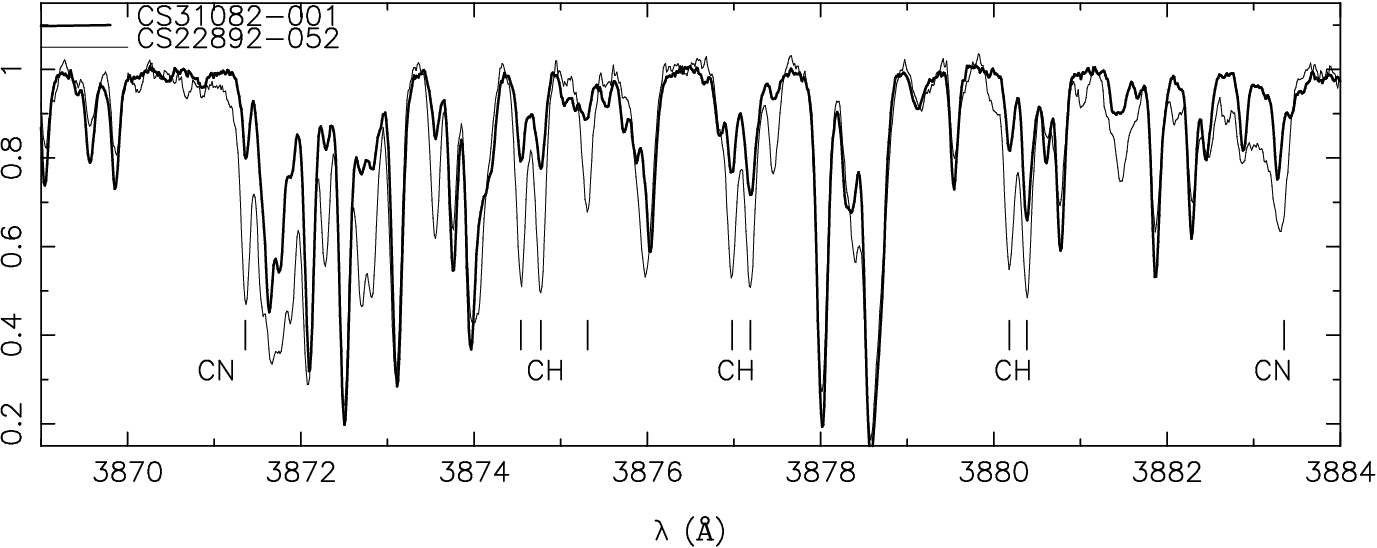}
    \includegraphics[angle=-90,width=17cm]{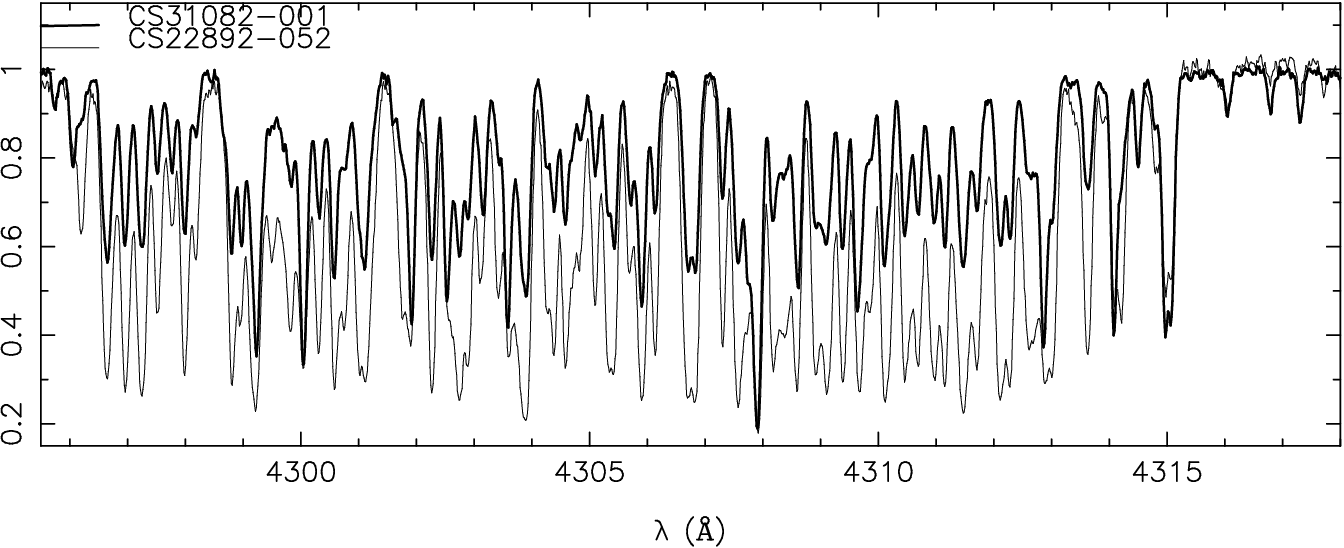}}
      \caption{Comparison of the spectrum of \cs (thick line) and \sned\,
    (narrow line) showing the much reduced molecular band strength in
    the former. {\it Upper panel:} CN 3871 and 3883\AA\ band heads;
    {\it lower panel:} G band (CH). The spectrum of \sned was obtained
    during the comissioning of the spectrograph UVES, and is publicly 
available(http://www.eso.org/science/uves\_comm/).} \label{F-CN}
   \end{figure*}

During this program, the bright (V = 11.7), very metal-poor halo giant \cs,
discovered during medium-resolution spectroscopic follow-up of candidate
metal-poor stars from the HK survey of Beers and colleagues (Beers, Preston, \&
Shectman \cite{beers92}; Beers \cite{beers99}; Beers et al. \cite{beers02b})
was found to be strongly enhanced in heavy neutron-capture elements.  Its large
$r$-element excess, low carbon and nitrogen content (reducing the CN and CH
molecular band contamination, see Table \ref{T-elements} and Fig.
\ref{F-CN}), and our superb VLT/UVES spectra permitted the first $^{238}$U
abundance measurement in a halo star (Cayrel et al. \cite{cayrel01}), as well
as a very accurate abundance of $^{232}$Th. This discovery opens the way to a
new, far more accurate and reliable radioactive chronometer for old stars,
based on the ratio U/Th rather than the conventional Th/Eu ratio, provided that
the initial production ratio of U/Th can be reliably estimated.

Uranium was first detected outside the Solar System in the spectra of
a handful of CP stars (Jaschek \& Malaroda \cite{jm70}; Cowley, Aikman
\& Fisher  
\cite{cow77} and references therein). The huge enhancements of the rare earths 
and other heavy elements in Ap stars are not fully understood, but
radiative support (e.g, Michaud et al. \cite{michaud76}) 
is generally favoured over a 
nucleosynthetic origin (Burbidge \& Burbidge \cite{burbige55}). Hence, the U 
abundance in CP stars is not relevant to galactic nucleosynthesis and 
radioactive chronometry and is not considered further here. 

Radioactive age determinations for halo stars have so far relied on the
hypothesis that the $r$-process pattern in such stars matches the Solar
pattern, as has been found in the few known $r$-process-enhanced extreme halo
stars \sned (Sneden et al. \cite{sneden96}, \cite{sneden00}) and \hd115 (Westin
et  al. \cite{westin00}). Indeed, it was tentatively concluded that a universal
production pattern exists for the $r$-process elements, independent of the
nature of the progenitor star(s) and the detailed history of the subsequent
formation of the stars we now observe.

Our first study of the {\it r}-process elements in \cs (Hill et al.
\cite{hill01}) suggested that significant deviations from the Solar
$r$-process pattern occur in this star. This paper substantiates this result
through a detailed abundance study of the star, taking advantage of the very
important recent progress in the measurement of the oscillator strengths of the
U~II  3859.57 \AA\ line (Nilsson et al. \cite{nilsson01a}), and the Th~II lines
(Nilsson et al. \cite{nilsson01b}). As a consequence of this new analysis we
have revised our measured value of the abundance ratio U/Th in \cs, and
simultaneously reduced its uncertainty, from derived observational errors on
the order of 0.15 dex to 0.11 dex. Moreover, new studies of the stellar
production of the actinides (Goriely \& Arnould \cite{ga01}) enable a 
significantly improved discussion of the age determination for \cs.

This paper is organized as follows:  Section \ref{observations} describes the
observations and data reduction, while section 3 presents a detailed
analysis of the elements up to nickel. Section  \ref{neutron capture} 
presents a re-analysis of the $r$-process elements, while section 
\ref{r universal} addresses the universality or otherwise of the 
$r$-process. In section \ref{age} we consider anew the age determination 
for \cs. Section \ref{conclusion} summarizes our conclusions.

\section{Observations and reductions} \label{observations}

\cs was first observed at high dispersion (R$>$45,000) with the new ESO-VLT
spectrograph UVES (Dekker et al. \cite{dekker00}) in August 2000, and was
immediately found to display outstandingly strong europium and thorium lines.
The star was promptly re-observed at the highest possible resolution
($R=75,000$ in the blue arm of the spectrograph) during a subsequent night, and
again in October 2000. Most of the spectra were obtained simultaneously in both
the blue and red arms of UVES, using a dichroic mirror, as summarized in Table
\ref{T-log}. 

\begin{table*}[htbp]
\caption{Log of the observations, and measured signal-to-noise ratio per
pixel from the extracted and co-added spectra}\label{T-log}
\begin{tabular}{llcccccc}
\hline
Date & MJD & Exp. time & $\lambda_{start} - \lambda_{end}$ & Resolving Power &
$\lambda_{start} -  \lambda_{end}$ &  Resolving Power &  V$_{r\: barycentric}$\\
& (s) & \multispan2 \hfill Blue \hfill & \multispan2 \hfill
Red \hfill & \\
2000/08/06 & 51763.396 & 1,200 & 335-456nm & 45,000 & 480-680nm & 45,000  & 139.03\\ 
2000/08/06 & 51763.412 & 1,200 & 335-456nm & 45,000 & 670-900nm & 45,000  &\\ 
2000/08/10 & 51767.281 & 3,000 & 335-456nm$^{*}$ & 70000 & --        & --  &\\ 
2000/08/10 & 51767.317 & 3,000 & 335-456nm$^{*}$ & 70000 & --        & --  &\\ 
2000/10/11 & 51829.169 & 7,200 & 310-390nm & 77,000 & --        & --  &\\
2000/10/14 & 51832.176 & 3,600 & 310-390nm & 77,000 & 480-680nm & 86,000  & 139.10\\ 
2000/10/14 & 51832.219 & 3,600 & 310-390nm & 77,000 & 480-680nm & 86,000  & 139.14\\ 
2000/10/15 & 51833.223 & 3,600 & 310-390nm & 77,000 & 480-680nm & 86,000  & 139.08\\ 
2000/10/17 & 51835.147 & 3,600 & 380-510nm & 75,000 & 680-1000nm & 86,000 & \\ 
2000/10/19 & 51837.184 & 3,600 & 380-510nm & 75,000 & 680-1000nm & 86,000 &\\ 
\end{tabular}

$^{*}$: obtained with Image Slicer

\begin{tabular}{lcrrrrrrr}
\hline
 Setting& Total &\multispan7 \hfill S/N per pixel \hfill\\
     & Exp. time (s)     &@310nm & @350nm& @390nm& @420nm& @490nm& @630nm & 
     @800nm\\
 310-390nm  &18,000 &35 &150 &350 \\
 335-460nm  & 6,000 &   &110 &250 &300\\
 380-510nm  & 7,200 &   &    &250 &250 &200\\
 480-680nm  &10,800 &   &    &    &   &250 & 340\\
 670-1000nm & 7,200 &   &    &    &   &    &     & 110 \\
\hline \hline
\end{tabular}
\end{table*}

Reductions were performed using the UVES context within MIDAS. The succession
of tasks included bias subtraction from all images, fit and subtraction of the
inter-order background from the object and flat-field images, and the averaging
of ten flat-fields per night into a master frame. The echelle orders of the
object were optimally extracted\footnote{In the two frames that were
obtained through the image slicer, the extraction was simply an average of all
pixels in the object profile, spread over four slices.} (assuming a Gaussian 
profile for the object perpendicular to the dispersion, and a
constant for the sky background); the same weights were applied to extract
the flat-field, which was then used to correct both for pixel-to-pixel
variations and the blaze of the instrument.  The wavelength calibration was
performed on Th-Ar lamp frames and applied to the extracted object. The
spectra were finally resampled to a constant wavelength step and normalized to
unity by fitting a spline function.  All spectra obtained with identical
spectrograph settings (same cross-disperser and same central wavelength) were
then co-added after radial-velocity correction.  The lower part of Table
\ref{T-log} summarizes the obtained signal-to-noise ratio per pixel (1
pixel$\sim$0.0015nm) at wavelengths of interest.

Radial velocities given in the table are only those computed from
the 480-580nm red spectra (which permitted corrections for small instrumental
shifts from the telluric absorption lines). In addition to the
spectra reported here, we took exposures of \cs with UVES again in
September 2001 (5-9 september) and the radial velocity was still
impressively unchanged (mean of four measurements in 2001 V$_{r\:
barycentric}=139.05\pm0.05 kms^{-1}$) and  Aoki (2002, private
communication) found V$_{r\:barycentric}=138.9\pm0.24 kms^{-1}$ from 
observations with HDS on the Subaru telescope. There is therefore {\it no
sign} of radial velocity variations in this star, and hence {\it no hint
that \cs  could be a binary}.

\section {Abundance analysis, and derived abundances of the lighter elements}

\begin{figure}
   \centering
   {\includegraphics[angle=0,width=8cm]{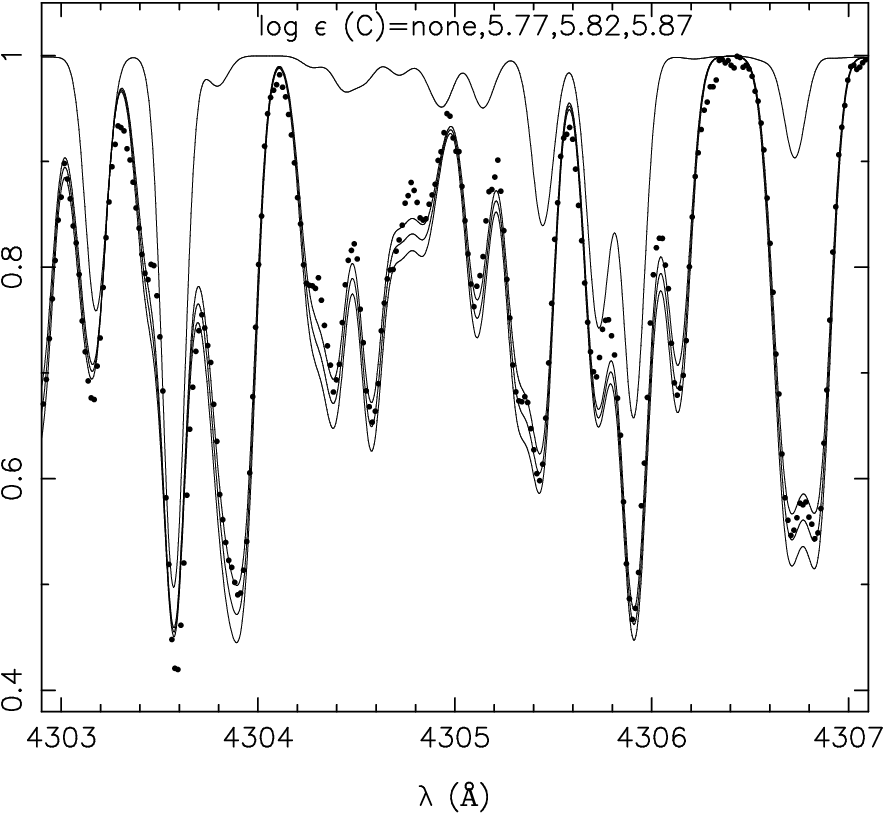}}
      \caption{Fit of CH lines of the G band in \cs. {\it Dots:} 
      observations: {\it lines:} synthetic spectra computed for the 
      abundances indicated in the figure.}
         \label{F-C}
   \end{figure}

\begin{figure}
   \centering
   {\includegraphics[angle=0,width=8cm]{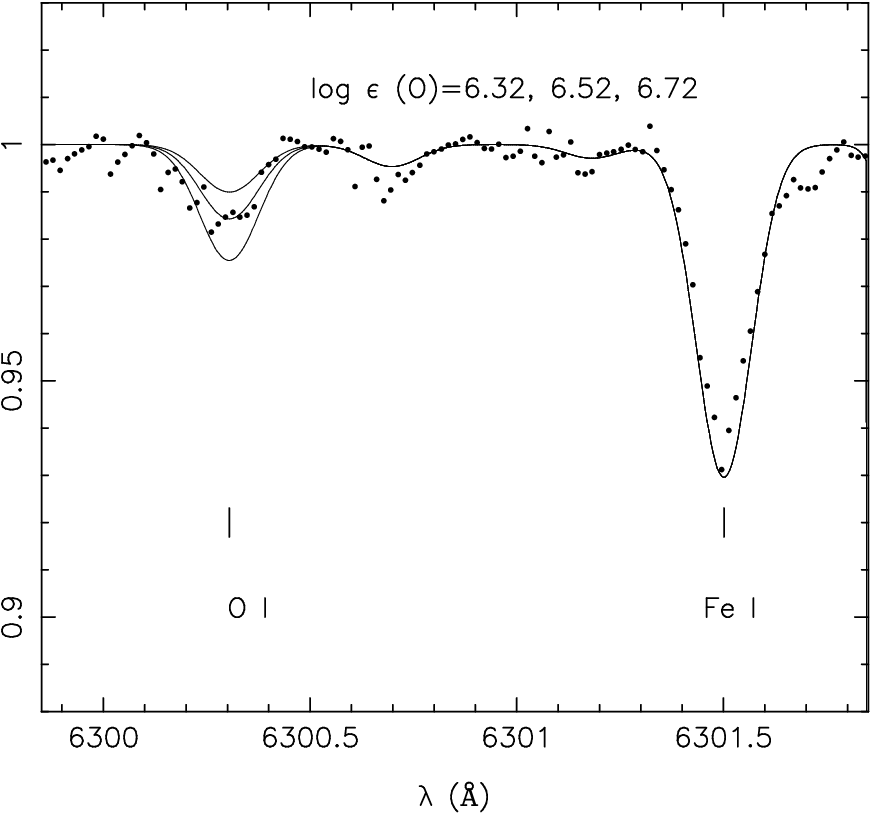}}
      \caption{The forbidden [O I] 6300 \AA\ line in \cs. Symbols as in
      Fig. \ref{F-C}}
         \label{F-O}
   \end{figure}

\subsection {Model atmosphere and stellar parameters}

The adopted model atmospheres (OSMARCS) were computed with the latest version
of the MARCS code, initially developed by Gustafsson et al. (\cite{gben75}) 
and subsequently updated by Plez et al. (\cite{pbn92}), Edvardsson et al.
(\cite{edvardsson}), and Asplund et al. (\cite{agke97}). The current version
includes up-to-date continuum and line opacities for atomic and molecular
species, treated in opacity sampling with more than 10$^5$ sampling points
between 910\AA\ and 20$\mu$m.  Models for \cs were computed for a
metallicity 1/1000th Solar, with the $\alpha$-elements boosted by 0.4~dex
relative to iron.

In our preliminary analysis of the star (Hill et al. \cite{hill01}),
we were using the synthetic spectrum code of Spite (\cite{codespite}
 and subsequent improvements in the last thirty years). 
In the present analysis we have employed a
more consistent approach based on the {\it turbospec} synthesis code
developed by Plez (Plez et al. \cite{codeplez}), which shares routines
and input data with OSMARCS. The latest version (Alvarez \& Plez
\cite{codeplez98}) features: Full chemical equilibrium including 
92 atomic and over 500 molecular species, Van der Waals collisional 
broadening by H, He, and H$_2$ following Anstee and O'Mara (\cite{anstee}), 
Barklem and O'Mara (\cite{barklem1}), and Barklem et al. (\cite{barklem2}),
and updated continuum opacities, and plane-parallel or spherical 
geometry.
The main differences between the Spite et al. and the Plez codes lie
in the continuum opacity, the source function (diffusion 
is included in the latter), and the collisional broadening calculation.

The effective temperature for the star was computed from multicolor
information, using the Alonso et al. (\cite{alonso99}) color-temperature
transformations.  A number of photometric data are available for \cs:
$UBVR_CI_C$ (subscript $C$ indicating the Cousins system) 
are from Beers et al. (\cite{beers02a}); the $V$ magnitude and Str\"omgren
photometry are from Twarog et al. (\cite{twarog00}); infrared data are
available from the DENIS (Fouqu\'e et al. \cite{fou00}) and 2Mass surveys
(Cutri et al. \cite{2mass}). A summary of these photometric data, and the
corresponding derived temperatures, is given in Table \ref{T-color}

\begin{table}[htbp]
\caption{Colors and effective temperature of \cs.}\label{T-color}
\begin{tabular}{llll}
\hline
 Index    & value &   T$_{eff}$ (K) & T$_{eff}$ (K)\\
& &E$(B-V)$& E$(B-V)$\\

& & =0.00  & =0.03 \\
 $V$     & 11.674$\pm$0.009& \\
 $(B-V)$ & 0.772$\pm$0.015 & 4822$\pm$120 & 4903$\pm$120 \\
 $b-y$   & 0.542$\pm$0.009 & 4917$\pm$70  & 4980$\pm$70 \\
 $(V-R)_{C}$ & 0.471$\pm$0.015 & 4842$\pm$150 & 5027$\pm$150 \\
 $(V-I)_{C}$ & 0.957$\pm$0.013 & 4818$\pm$125 &  4987$\pm$125 \\
 $(V-K)$ & 2.232$\pm$0.008 & 4851$\pm$50 & 4967$\pm$50 \\
\hline\hline
\end{tabular}
\end{table}

At a Galactic latitude $l = -76\degr$, the observed colors of \cs are not
expected to be significantly affected by reddening.  The Burstein \& Heiles
(\cite{bh82}) maps suggest negligible reddening; the Schlegel et al.
maps (Schlegel, Finkbeiner, \& Davis \cite{sc98}) suggest E(B-V)$\approx 0.03$.
Table \ref{T-color} lists effective temperatures derived both for the situation
of no reddening, and for an adopted reddening of 0.03 mags.  The no-reddening
values are in better agreement with the derived excitation temperature from Fe~I
lines.  The final adopted temperature of T$_{eff}$=4825~K is consistent with
that obtained from the excitation equilibrium of the Fe~I lines. A gravity of
$\log g$=1.5$\pm$0.3 dex was assumed in order to satisfy the ionization
equilibrium of iron and titanium, and a microturbulence velocity of
$\xi$=1.8$\pm$0.2 km.s$^{-1}$ was obtained from the requirement that strong and
weak lines of iron yield the same abundance.

In this paper we are mostly concerned with the relative abundances of elements 
in this star, especially the abundance pattern of the heavy neutron-capture
elements.  The relative abundances are only very weakly dependent on the
adopted stellar parameters; all of the lines of interest respond similarly to
small changes in temperature and gravity, hence the pattern of the heavy
elements relative to one another is hardly affected.  A thorough discussion of
errors is provided in section \ref{errors}.

\begin{figure*}
   \centering
   {\includegraphics[angle=0,width=5.9cm]{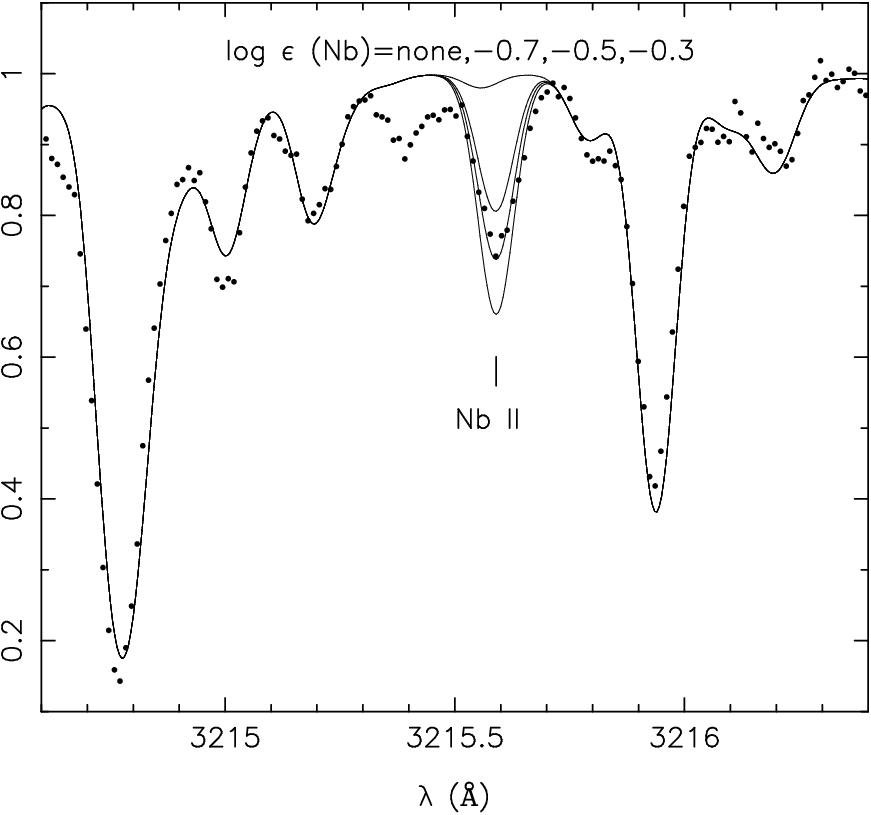}
    \includegraphics[angle=0,width=5.9cm]{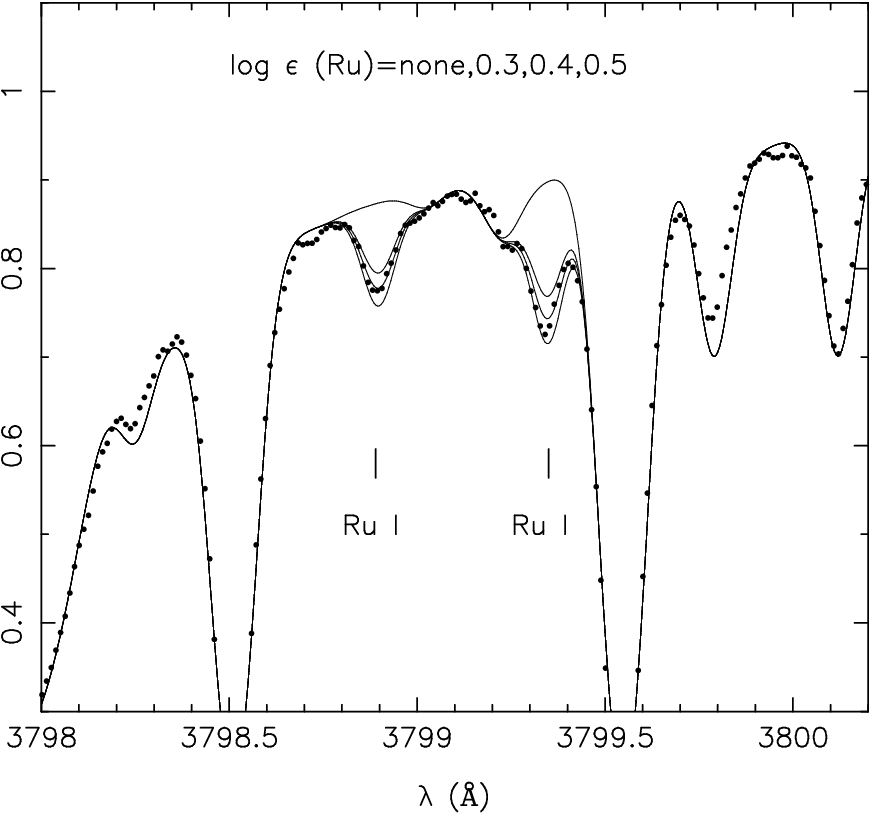}
    \includegraphics[angle=0,width=5.9cm]{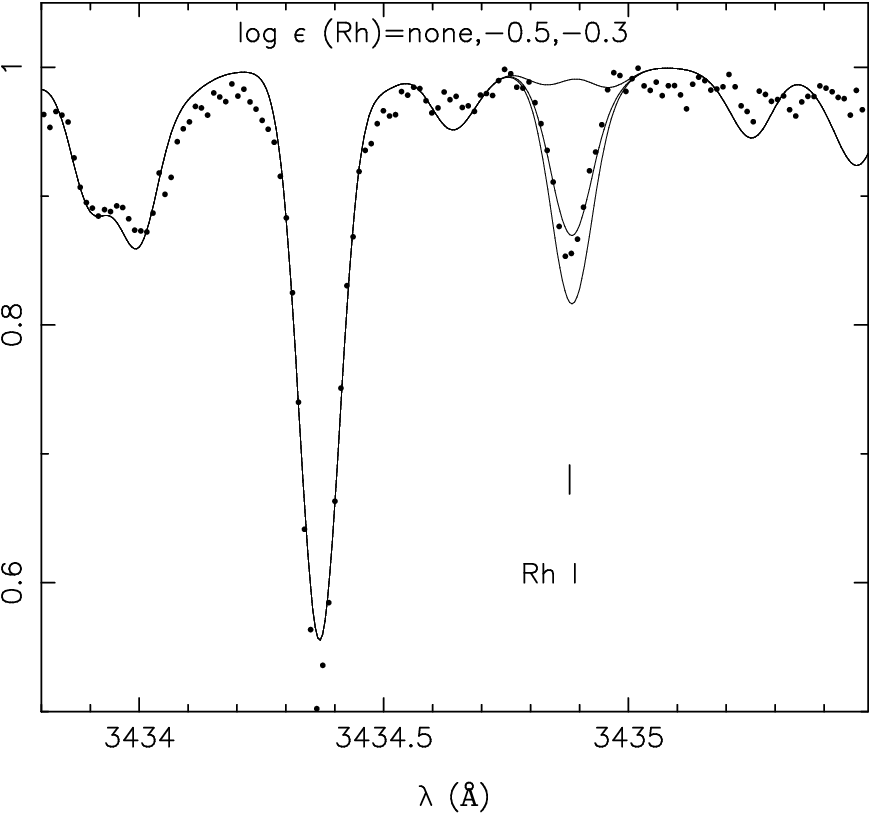}}
      \caption{The observed Nb II 3215 \AA, Ru I 3799 \AA, and Rh I
      3434 \AA\ lines in \cs. Symbols as in
      Fig. \ref{F-C}.}
         \label{F-Nb-Ru-Rh}
   \end{figure*}

   \begin{figure*}
   \centering
   {\includegraphics[angle=0,width=5.9cm]{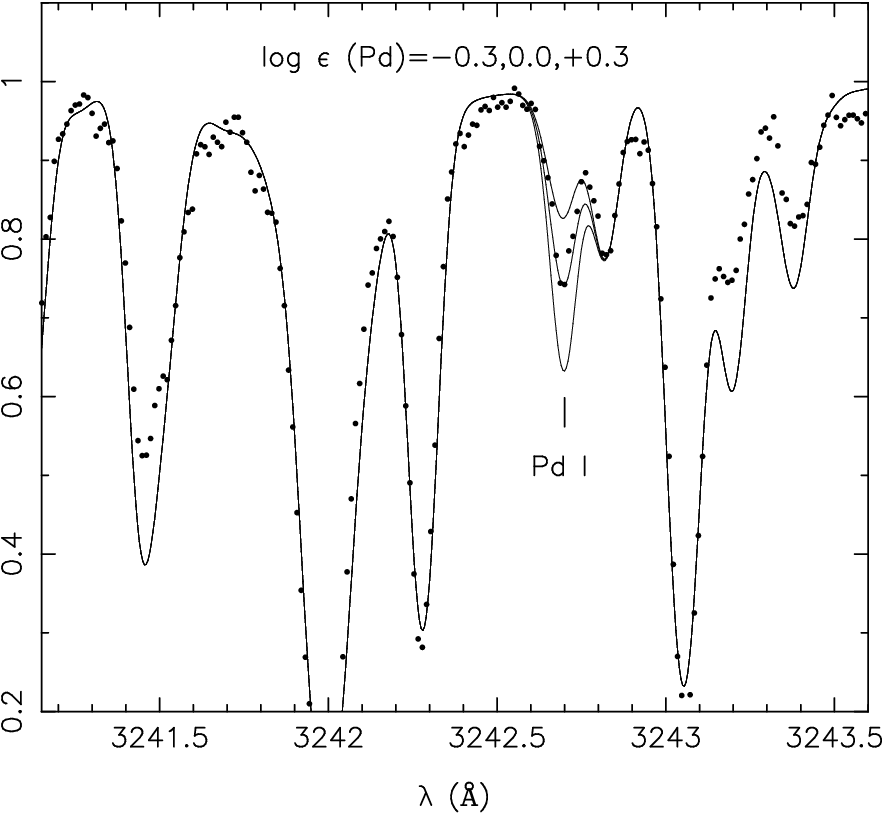}
    \includegraphics[angle=0,width=5.9cm]{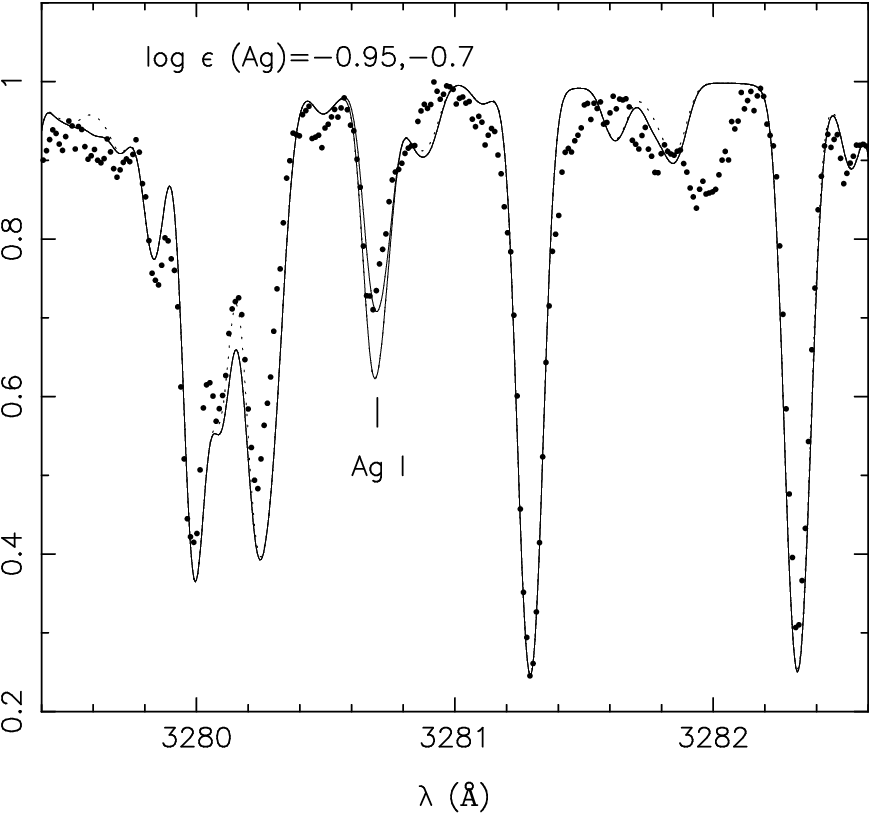}
    \includegraphics[angle=0,width=5.9cm]{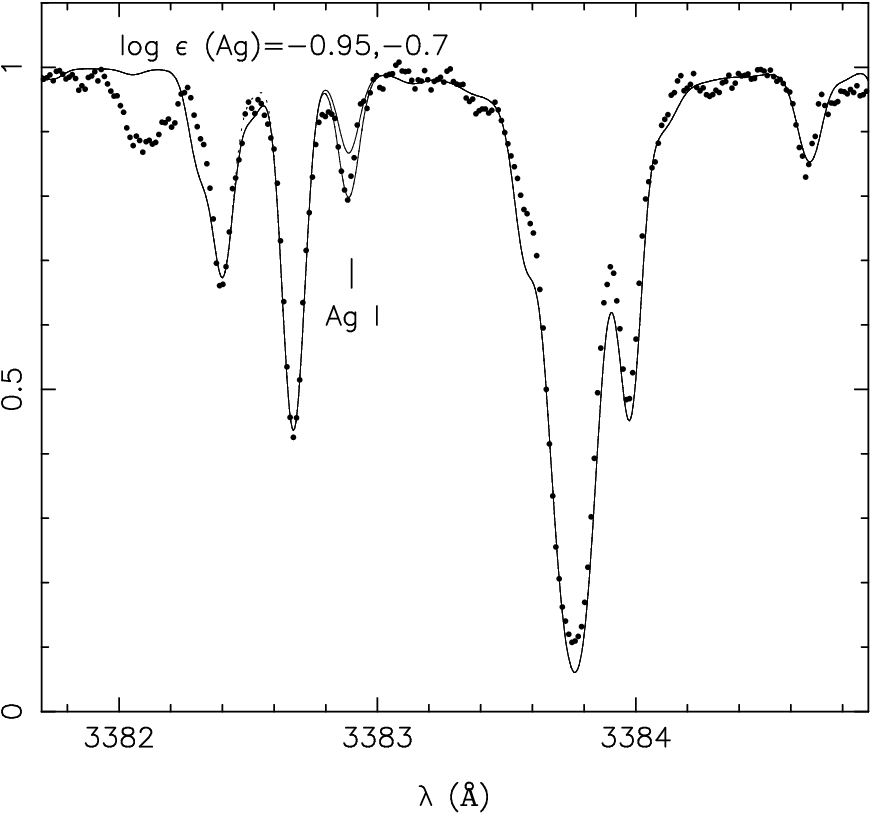}}
      \caption{The observed Pd I 3242 \AA, Ag I 3281 \AA, and 3383 \AA\
    lines in \cs. Symbols as in
      Fig. \ref{F-C}.}
         \label{F-Pd-Ag}
   \end{figure*}

\subsection{Abundances of light and iron-peak elements} \label{light}

Most of the abundances for the light and the iron-group
elements were determined via equivalent width measurement of a
selection of unblended lines. Exceptions are Li, C, N, and O, for which
synthesis spectra were directly compared to the observed spectrum.
The linelist used for all light and iron-peak elements will be
published together with the analysis of the complete sample of our
Large Program. For the compilation of this linelist, we used the VALD2
compilation of Kupka et al. \cite{vald2}.

Table \ref{T-elements} lists the mean abundances\footnote{In the classical 
notation where $\log \epsilon$(H) = 12 and $\rm [X/H]=\log (N_{X}/N_{H})_{*} - 
\log(N_{X}/N_{H})_{\sun}$.}, dispersion of the single line measurements 
around the mean ($\sigma$), and the number of lines used to determine 
the mean abundances of all measured elements from lithium to zinc.  Also
listed are the abundances relative to iron, [X/Fe], and the total uncertainty
on this ratio, $\Delta [X/Fe]$, including errors linked both to observations
and the choice of stellar parameters (similarly to the $\Delta [X/Th]$ 
reported in Table \ref{T-error} and explained in section \ref{errors}).
Notes on specific elements are:

\begin{table}
\begin{center}
\caption{LTE abundances for lighter elements in CS\,31082-001
\label{T-elements} }
\begin{tabular}{llccc@{ }rc@{ }c}
\hline
El.   & Z& $\log \epsilon$ & [X/H] & $\sigma$ & N$_{lines}$ & [X/Fe] & 
$\Delta$[X/Fe]\\
\multispan2 $^{12}C/^{13}C$ & $>$20  \\
Li I  & 3 & 0.85 &   &  & 1 & & 0.11\\
C     & 6 & 5.82 & $-$2.7   & 0.05  & &  +0.2 &  \\ 
N     & 7 & $<$5.22 &$<-$2.7 &   &  &$<$+0.2&     \\ 
O I   & 8 & 6.52 &$-$2.31 &      &  1 &  0.59 & 0.20 \\
Na I  &11 & 3.70 &$-$2.63 & 0.02 &  2 &  0.27 & 0.13 \\
Mg I  &12 & 5.04 &$-$2.54 & 0.13 &  7 &  0.36 & 0.16 \\ 
Al I  &13 & 2.83 &$-$3.64 &      &  1 & $-$0.74 & 0.17 \\
Si I  &14 & 4.89 &$-$2.66 &      &  1 &  0.24 & 0.10 \\
K I   &19 & 2.87 &$-$2.25 &0.08  & 2  &  0.65 & 0.10 \\
Ca I  &20 & 3.87 &$-$2.49 & 0.11 & 15 &  0.41 & 0.08 \\
Sc II &21 & 0.28 &$-$2.89 & 0.07 &  7 &  0.01 & 0.06 \\
Ti I  &22 & 2.37 &$-$2.65 & 0.09 & 14 &  0.25 & 0.10 \\ 
Ti II &22 & 2.43 &$-$2.59 & 0.14 & 28 &  0.31 & 0.05 \\ 
Cr I  &24 & 2.43 &$-$3.24 & 0.11 &  7 & $-$0.34 & 0.11 \\
Mn I  &25 & 2.14 &$-$3.25 & 0.09 &  6 & $-$0.35 & 0.10 \\
Fe I  &26 & 4.60 &$-$2.90 & 0.13 &120 &  0.00 &  \\ 
Fe II &26 & 4.58 &$-$2.92 & 0.11 & 18 &  0.02 &  \\ 
Co I  &27 & 2.28 &$-$2.64 & 0.11 &  4 &  0.26 & 0.12 \\
Ni I  &28 & 3.37 &$-$2.88 & 0.02 &  3 &  0.02 & 0.14 \\
Zn I  &30 & 1.88 &$-$2.72 & 0.00 &  2 &  0.18 & 0.09 \\
\hline \hline
\end{tabular}
\end{center}
\end{table}

\noindent{\it Lithium}\\
With an equivalent width of more than 15 m\AA, the lithium 6708 \AA\ line is 
easily detected in this star. The abundance was determined using spectrum
synthesis techniques to account for the doublet nature of the line.  With an
abundance of $\log \epsilon$(Li)=0.85, \cs\, falls well below the lithium 
plateau for hot dwarf halo stars, as expected for a red giant. However,
not all lithium in this star has been diluted after the first dredge-up, as
observed in many metal-poor giants. In fact, in a sample of 19 giants with
[Fe/H] $\leq -2.7$ (Depagne et al. 2002), we find that among the
eight giants with gravities close to $\log g$=1.5, four have detectable
lithium.  Hence, in this respect, \cs is not exceptional.

\noindent{\it Carbon and nitrogen}\\
These two elements are detected via the molecular bands of CH and CN.  Line
lists for $^{12}$CH, $^{13}$CH, $^{12}$C$^{14}$N, and $^{13}$C$^{14}$N were
included in the synthesis.  The CN linelists were prepared in a similar fashion
 as the TiO linelist of Plez (\cite{pleztio}), using data from Cerny et
al. (\cite{cerny}), Kotlar et al. (\cite{kotlar}), Larsson et al.
(\cite{larsson}), Bauschlicher et al (\cite{bausch}), Ito et al. (\cite{ito}),
Prasad \& Bernath (\cite{prasada}), Prasad et al. (\cite{prasadb}), and Rehfuss
et al. (\cite{rehfuss}).  Programs by Kotlar were used to compute wavenumbers
of transitions in the red bands studied by Kotlar et al. (\cite{kotlar}).  For
CH, the LIFBASE program of Luque and Crosley (\cite{lifbase}) was used to
compute line positions and gf-values. Excitation energies and isotopic shifts
(LIFBASE provides only line positions for $^{12}$CH) were taken from the line
list of J\o rgensen et al. (\cite{jorgensen}). By following this procedure, a
good fit of CH lines could be obtained, with the exception of a very few
lines which we removed from the list.

$^{13}$C isotopic lines could not be detected, hence we only provide a lower 
limit on the $^{12}$C/$^{13}$C ratio, based on the non-detection of $^{13}$CH 
lines.

The carbon abundance was derived primarily from CH lines in the region
4290-4310 \AA\ (the CH A-X 0-0 bandhead), which is almost free from intervening
atomic lines. We derive an abundance $\log \epsilon (C)= 5.82 \; \pm
\;0.05$ (see Fig. \ref{F-C}).
With this same abundance, a good fit is obtained in the more crowded regions
around 3900 \AA\  and 3150 \AA, where the B-X and C-X bandheads occur.

The nitrogen abundance was then derived from CN lines. It was only possible to
set an upper limit, since the CN lines are extremely weak. A conservative estimate
of the nitrogen abundance is $\log \epsilon (N)= 5.22$, from the 3875-3900 \AA\
B-X 0-0 bandhead. An attempt was also made to use the NH 0-0 bandhead of the A-X
system around 3350 \AA. Lines were extracted from the Kurucz line database
(\cite{kuruczcd}), and the gf-values were scaled with the mean correction
derived by comparison of the Kurucz and Meyer \& Roth (\cite{meyer}) gf-values
for the 2 (0-0) R$_1$(0) and $^R$Q$_{21}$(0) lines at 3358.0525 \AA\  and
3353.9235 \AA. This derived correction, $-0.807$ in log(gf), was applied to all
of the NH lines.  The fit was poor, but if the gf-values are correct,
which is a bold assumption, the nitrogen abundance is at most $\log \epsilon
(N)= 5.02$. Given the many uncertainties attached to this latter determination,
we adopt the more conservative estimate obtained from CN.

\noindent{\it Oxygen}\\
The forbidden oxygen line at 630nm is clearly detected in the three spectra
taken in October 2000, where the motion of the Earth relative to the star
resulted in a Doppler shift that moved this weak line clear of the 
neighboring telluric absorption lines (Fig. \ref{F-O}). The measured
equivalent width of the [OI] line is 2.7 m\AA; the corresponding 
abundance is $\log \epsilon$(O)=6.52, which in turn 
implies an overabundance of oxygen with respect
to iron of [O/Fe]=+0.59$\pm$0.2.  

Thus, \cs has an oxygen abundance that
is consistent with the mild oxygen enhancement observed for other halo stars,
at least when derived from the same forbidden [OI] line (e.g.  Barbuy et al.
\cite{barbuy88}; Sneden et al. \cite{sneden01}; Nissen et al. \cite{nissen01}).
The linearly increasing trend of [O/Fe] with decreasing metallicity suggested
from measurements of OH lines in the UV of halo turnoff stars (e.g., Israelian
et al. \cite{israelian98}; Boesgaard et al. \cite{boesgaard99}) is not a
relevant comparison here, since it is known that there exist systematic
differences between the two indicators (UV OH and [OI]), that could arise, for
example, from temperature inhomogeneity effects (see Asplund \& Garc\'ia
P\'erez \cite{asplund01}).

All of the $\alpha$-elements in \cs are enhanced by 0.35 to 0.6 dex
relative to iron ([$\alpha$/Fe]=+0.37$\pm$0.13, 
where $\alpha$ is the mean of O, Mg, Si, Ca, and Ti),
consistent with the observed behavior of other metal-poor halo stars.

Potassium is also observed in \cs ,  from the red lines at 7664 \AA\  and
7698 \AA; an LTE analysis yields [K/Fe]$_{LTE}$=+0.65. However, Ivanova \&
Shimanskii (\cite{ivanova00}) have shown that these transitions suffer from
significant NLTE effects.  For a star with T$_{eff} = 4800$K, $\log g$=1.5, 
the correction amounts to $\sim -0.33$ dex. Therefore, the true potassium 
abundance in \cs\, should be around [K/Fe]=+0.3 dex.

\noindent{\it Iron group (Cr through Ni)}\\
Among the elements of this group, all are depleted to a similar level as iron,
although Co is up by $\sim$0.3 dex, while Cr and Mn are down 
by $\sim$0.3 dex, a typical
behaviour for metal-poor stars (e.g., McWilliam et al. \cite{mcwilliam95};
Ryan, Norris, \& Beers \cite{ryan96}). Thus, both in this respect, and from 
its standard [$\alpha$/Fe] enhancement, \cs appears to be a typical very
metal-poor halo star, except for its high neutron-capture-element abundances.
This behavior is understandable if the light elements are produced in the
hydrostatic burning phase of the evolution of massive stars, whereas the
r-process elements are produced during the explosive synthesis phase of the
SNe.

\noindent{\it Zinc}\\
Zinc is present in \cs at a level comparable to the iron-group elements,
and thus does not share the strong enhancement of neutron-capture elements
in this star. This is relevant to the ongoing discussions on the
origin of Zn in halo stars (Umeda \& Nomoto \cite{umeda02} and
references therein), and the importance of the neutron-capture channel
for Zn production.  Either Zn does not arise from neutron-capture processes, or
these processes are unimportant in r-process conditions. 

\section{Neutron-capture elements} \label{neutron capture}

\begin{figure}
   \centering
   {\includegraphics[angle=0,width=4.2cm]{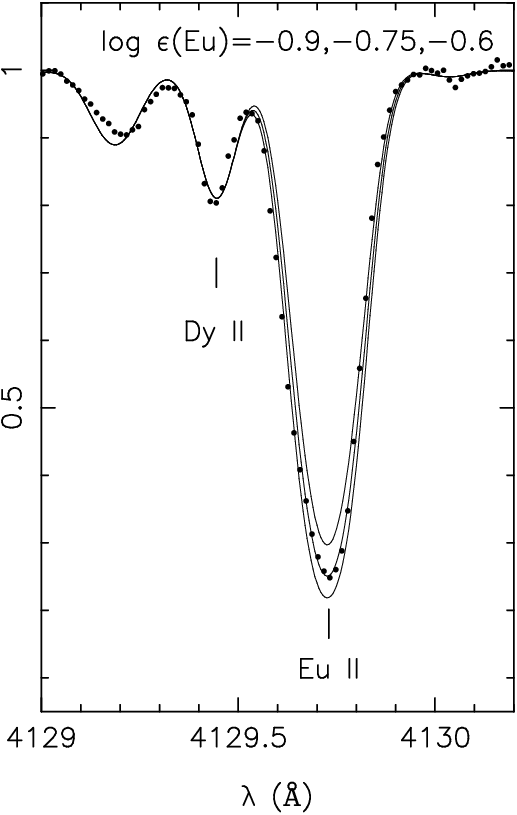}
    \includegraphics[angle=0,width=4.2cm]{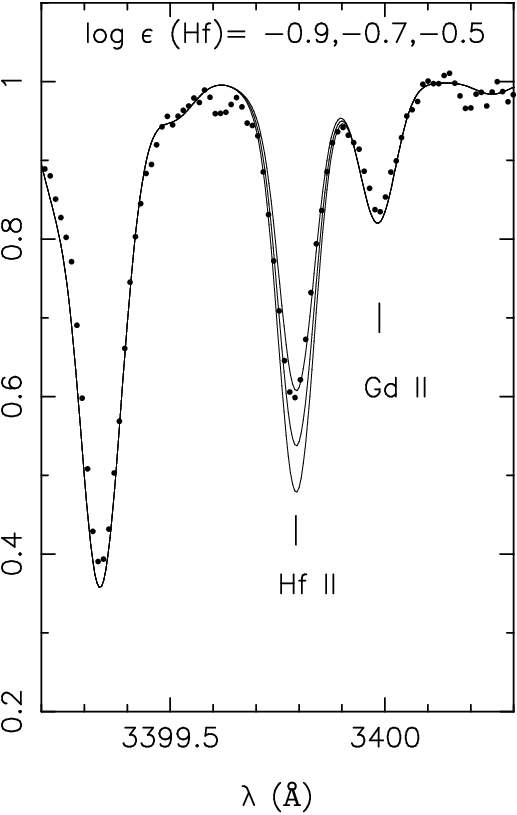}}
      \caption{The observed Eu 4129 and Hf II 3999 \AA\  line in \cs. Symbols as in
      Fig. \ref{F-C}.}
         \label{F-HfEu}
   \end{figure}

In this section we discuss our adopted linelist, and examine the possible
sources of error affecting the abundance determination of neutron-capture
elements.  We then explore the abundance patterns of elements in the three
r-process peaks. Eight elements with atomic numbers 38$\leq$Z$\leq$47 were
measured in the region of the  first peak.  The second peak is the best
constrained, with thirteen elements measured in the range 56$\leq$Z$\leq$72. 
The third peak and the actinides are probed by four elements with measured
abundances in the region 76$\leq$Z$\leq$92.

\subsection{Linelists and physical data}

\begin{figure*}
   \centering
   {\includegraphics[angle=0,height=7.1cm]{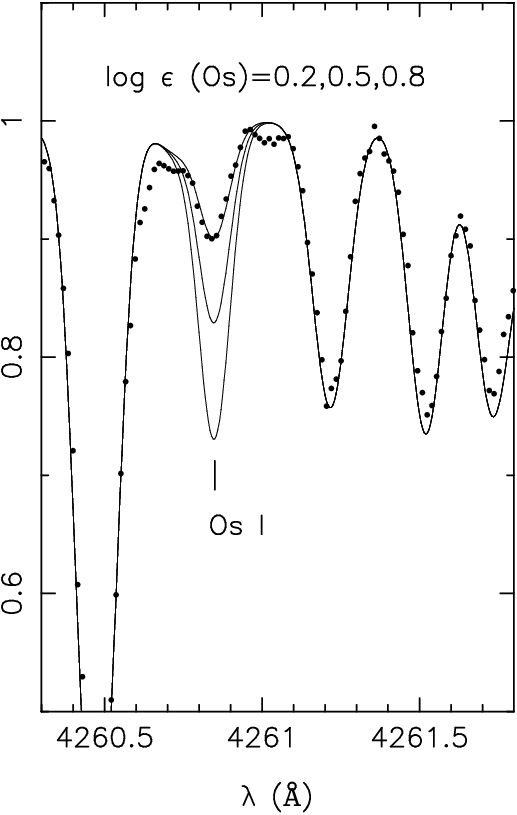}
    \includegraphics[angle=0,height=7.1cm]{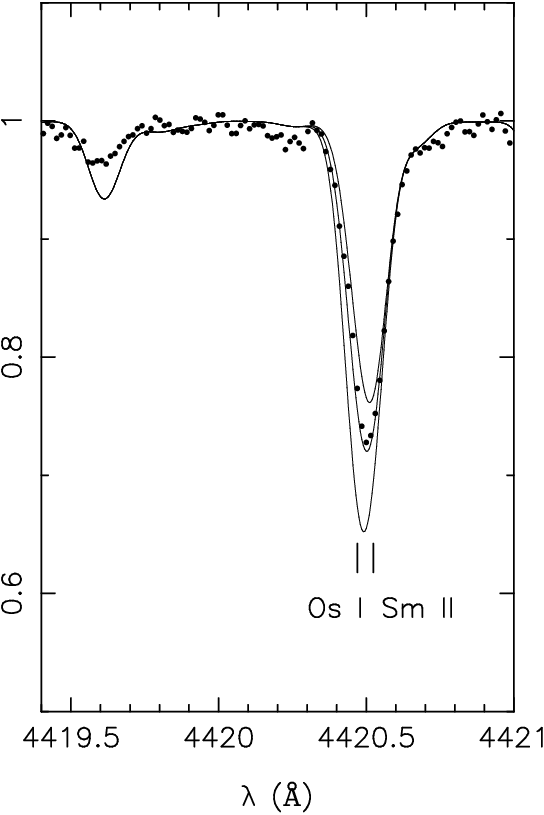}
    \includegraphics[angle=0,width=7.7cm]{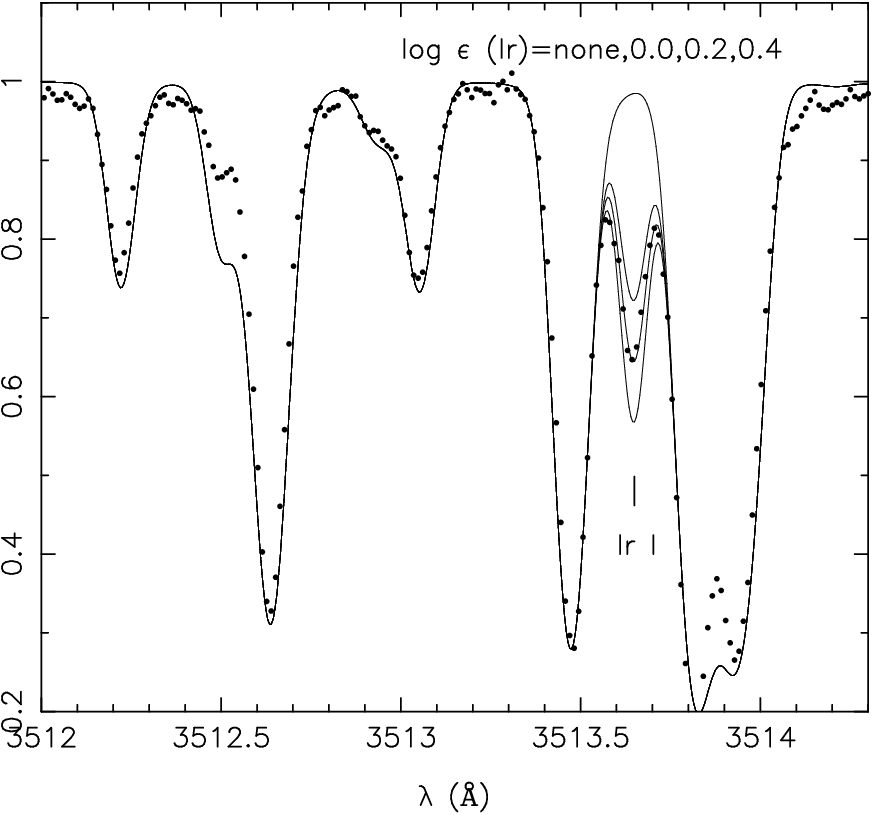}}
      \caption{The observed Os~I 4261 \AA, 4420 \AA, and Ir~I 3513 \AA\ 
    lines in \cs. Symbols as in
      Fig. \ref{F-C}.}
         \label{F-Os-Ir}
   \end{figure*}

\begin{figure}
   \centering
   {\includegraphics[angle=0,width=8cm]{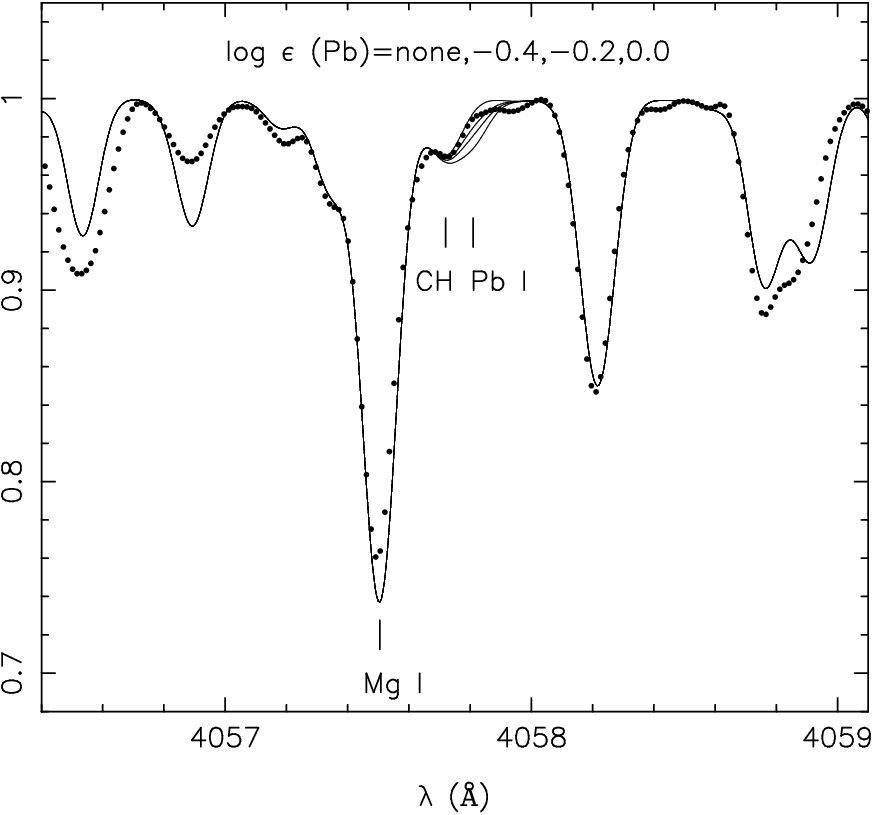}
   \includegraphics[angle=0,width=8cm]{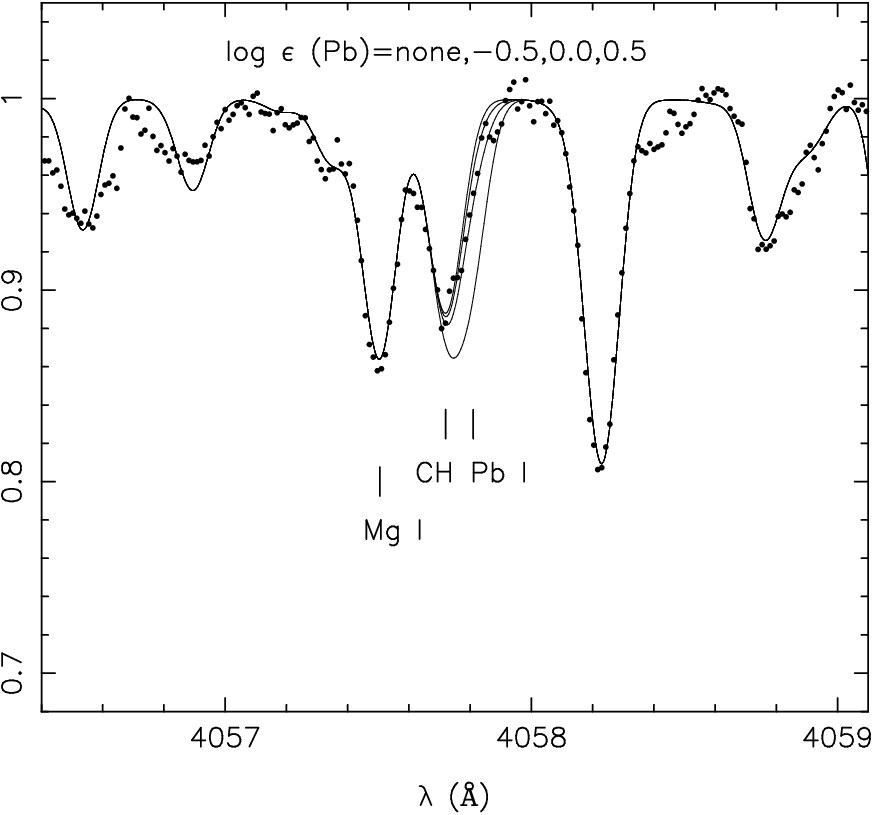}}
      \caption{The observed Pb I 4057 \AA\ lines in \cs 
      (upper panel) and \sned (lower panel). Symbols as in Fig. \ref{F-C}.} 
        \label{F-Pb}
   \end{figure}

Lines from 28 neutron-capture elements were observed in
\cs, mainly concentrated in the blue and UV parts of the spectrum. 
The full linelist for heavy elements is provided in an Appendix to this
paper, which also lists the references for our adopted oscillator strengths.
When available from the paper of Sneden et al. \cite{sneden96}, the same
oscillator strengths were adopted in order to make the comparison to the star
\sned\, easier. However, more lines were measured here, hence we had to
supplement this compilation with additional information. 
In particular, we note that new results on the
lifetimes and branching factors of both uranium (Nilsson et al.
\cite{nilsson01a}) and thorium (Nilsson et al. \cite{nilsson01b}) transitions
are now available; we make use of them here (Table \ref{T-Th}). The
Appendix also lists equivalent widths of the lines in \cs and the individual
derived abundances. In cases when blending was severe, or hyperfine structure
was important, abundances were determined by comparing the observations
directly to synthetic spectra (in these cases, no equivalent widths are listed
in the Appendix). Hyperfine structure was included for the Ba and Eu lines.  
The mean abundances obtained for each element are listed in Table
\ref{T-heavy}.  Figs. \ref{F-Nb-Ru-Rh} to \ref{F-U} are representative
examples of the quality of the observed spectrum, and the fits to synthetic
spectra. 

\begin{table}[htbp]
\caption{Th and U lines used in the abundance determination}\label{T-Th}
\begin{center}
\begin{tabular}{lccc}
\hline
$\lambda$(\AA) & $\chi_{ex}$(eV) & $\log gf$ & $\log \epsilon$\\
Th II \\
3351.229 &  0.188 & $-$0.600 &     $-$0.88 \\
3433.999 &  0.230 & $-$0.537 &     $-$0.80 \\
3435.977 &  0.000 & $-$0.670 &     $-$0.85 \\
3469.921 &  0.514 & $-$0.129 &     $-$0.93 \\
3675.567 &  0.188 & $-$0.840 &     $-$0.83 \\
4019.129 &  0.000 & $-$0.228 &     $-$1.04 \\
4086.521 &  0.000 & $-$0.929 &     $-$0.96 \\
4094.747 &  0.000 & $-$0.885 &     $-$1.05 \\
U II \\
3859.571 &  0.036 & $-$0.067 &     $-$1.92 \\
\hline\hline
\end{tabular}
\end{center}
\end{table}

\begin{table}[htbp]
\begin{center}
\caption{Neutron-capture-element abundances in CS\,31082-001
\label{T-heavy} }
\begin{tabular}{llcccrc}
\hline
El.& Z   &$\log \epsilon$(X)& $\sigma$& $\Delta \log \epsilon$
&N$_{lines}$ & [X/Fe]\\
   &     &                  &         & (X/Th) \\
Sr &38   & 0.72   &0.03   &  0.08 & 3 & +0.65\\ 
 Y &39   &$-$0.23   &0.07   &  0.06 & 9 & +0.43\\ 
Zr &40   & 0.43   &0.15   &  0.09 & 5 & +0.73\\ 
Nb &41   &$-$0.55   &       &  0.15 & 1 & +0.93\\ 
Ru &44   & 0.36   &0.10   &  0.14 & 5 & +1.42\\ 
Rh &45   &$-$0.42   &0.03   &  0.13 & 3 & +1.36\\ 
Pd &46   &$-$0.05   &0.10   &  0.15 & 3 & +1.16\\ 
Ag &47   &$-$0.81   &0.17   &  0.22 & 2 & +1.15\\ 
Ba &56   & 0.40   &0.17   &  0.11 & 6 & +1.17\\ 
La &57   &$-$0.60   &0.04   &  0.06 & 5 & +1.13\\ 
Ce &58   &$-$0.31   &0.10   &  0.04 & 9 & +1.01\\ 
Pr &59   &$-$0.86   &0.12   &  0.06 & 6 & +1.33\\ 
Nd &60   &$-$0.13   &0.17   &  0.05 &18 & +1.27\\ 
Sm &62   &$-$0.51   &0.16   &  0.06 & 9 & +1.38\\ 
Eu &63   &$-$0.76   &0.11   &  0.05 & 9 & +1.63\\ 
Gd &64   &$-$0.27   &0.15   &  0.06 & 9 & +1.51\\ 
Tb &65   &$-$1.26   &0.07   &  0.04 & 7 & +1.74\\ 
Dy &66   &$-$0.21   &0.13   &  0.07 & 6 & +1.55\\ 
Er &68   &$-$0.27   &0.08   &  0.09 & 5 & +1.70\\ 
Tm &69   &$-$1.24   &0.10   &  0.08 & 4 & +1.66\\ 
Hf &72   &$-$0.59   &       &  0.17 & 2 & +1.43\\ 
Os &76   & 0.43   &0.17   &  0.16 & 3 & +1.30\\ 
Ir &77   & 0.20   &       &  0.11 & 2 & +1.75\\ 
Pb &82   &$<-$0.2:&       &       & 1 & \\
Th &90   &$-$0.98   &0.05   &  0.11 & 8 & +1.83\\ 
 U &92   &$-$1.92   &       &  0.11 & 1 & +1.49\\ 
\hline \hline 
\end{tabular}
\end{center}
\end{table}

   \subsection{Error budget} \label{errors}

Table \ref{T-error} summarizes the various sources of uncertainties affecting
the derived neutron-capture-element abundances in \cs.  Stochastic errors
($\Delta$(obs) listed in column 6) arise from random uncertainties in
the oscillator strengths (gf values) and in the measured equivalent widths.
The magnitude of this error is estimated as $\sigma$/$\sqrt{N-1}$
(where $\sigma$ is the rms around the mean abundance) when N$\geq$2 lines of a
given element are observed, otherwise as the quadratic sum of the estimated
error on the adopted gf value and the fitting uncertainty. Systematic
uncertainties include those which exist in the adopted oscillator strengths,
in the equivalent width measurements, mostly related to continuum location, and
in the adopted stellar parameters. The first
is extremely difficult to assess and is not considered explicitly here,
although it might be significant 
(gf values from various sources may be cross-checked, but often 
only one source is available).  The second should be negligible,
given the very high quality of our data.  Hence we have examined here only
those errors linked to our choice of stellar parameters.  These were estimated
by varying T$_{eff}$ by +100K, $\log g$ by +0.3 dex, and $\xi$ by +0.2kms$^{-1}$
in the stellar atmosphere model (columns 2, 3, and 4, respectively).
The quantity $\Delta$(T,$\log g$,$\xi$) listed in column 5 is the total impact
of varying each of the three parameters, computed as the quadratic sum of
columns 2, 3, and 4. The total uncertainty $\Delta$(total) (column 7)
on the {\it absolute} abundance of each element ($\log \epsilon$(X))
is computed as the quadratic sum of the stochastic
and systematic errors.  Columns 8 and 9 list the total uncertainties
on the abundance ratios X/U and X/Th.  Note that, due to the similarity of the 
response of a given set of elements to changes in the stellar parameters, 
systematic errors largely cancel out in the measured {\it ratios} of these 
elements, reducing the uncertainty affecting the {\it relative}
abundances. However, for the few species that are determined from
neutral lines, the stellar parameters uncertainties impact on the
[X/Th] or [X/U] ratio is not negligible anymore (eg. Ru, Rh, Pd).

In the following discussion, since we are mostly concerned by the {\it
relative} abundance ratios, we chose to consider the total uncertainties on the
X/Th ratio as representative of the uncertainty on the abundance pattern (in
Figs. \ref{F-r-sun} and \ref{F-r-cs22}, and listed in column $\Delta \log
\epsilon$(X/Th) of table \ref{T-heavy}).

\begin{table*}
\begin{center}
\caption{Error budget for neutron-capture elements
\label{T-error} }
\begin{tabular}{lcccccccc}
\hline
El. & $\Delta$T &  $\Delta\log g$ &  $\Delta \xi$ & $\Delta$(T,$\log g$,$\xi$) 
& $\Delta$(obs) & $\Delta$(total) 
& $\Delta \log \epsilon$(X/U) & $\Delta \log \epsilon$(X/Th) \\
      & +100K & +0.3dex       & +0.2kms$^{-1}$ \\
(1) & (2) & (3) & (4) & (5) & (6) &(7) & (8) & (9) \\ 
\hline
Sr II & 0.064 & 0.045 & $-$0.047 & 0.092 & 0.030 & 0.10 & 0.13 &  0.08 \\ 
 Y II & 0.068 & 0.082 & $-$0.057 & 0.121 & 0.023 & 0.12 & 0.13 &  0.06 \\ 
Zr II & 0.066 & 0.087 & $-$0.034 & 0.114 & 0.075 & 0.14 & 0.14 &  0.09 \\ 
Nb II & 0.044 & 0.091 & $-$0.019 & 0.128 & 0.150 & 0.20 & 0.19 &  0.15 \\
Ru I  & 0.079 &$-$0.021 &$-$0.012 & 0.159 & 0.050 & 0.17 & 0.18 &  0.14 \\
Rh I  & 0.080 &$-$0.019 &$-$0.007 & 0.162 & 0.021 & 0.16 & 0.17 &  0.13 \\
Pd I  & 0.081 &$-$0.023 &$-$0.024 & 0.166 & 0.071 & 0.18 & 0.19 &  0.15 \\
Ag I  & 0.081 &$-$0.022 &$-$0.020 & 0.166 & 0.170& 0.24 & 0.24 &  0.22 \\
Ba II & 0.096 & 0.046 & $-$0.075 & 0.122 & 0.076 & 0.14 & 0.16 &  0.11 \\ 
La II & 0.080 & 0.083 & $-$0.056 & 0.128 & 0.020 & 0.13 & 0.12 &  0.06 \\ 
Ce II & 0.078 & 0.089 & $-$0.013 & 0.119 & 0.035 & 0.12 & 0.12 &  0.04 \\ 
Pr II & 0.078 & 0.089 & $-$0.012 & 0.119 & 0.054 & 0.13 & 0.12 &  0.06 \\ 
Nd II & 0.078 & 0.086 & $-$0.024 & 0.119 & 0.041 & 0.13 & 0.12 &  0.05 \\     
Sm II & 0.082 & 0.089 & $-$0.010 & 0.122 & 0.057 & 0.13 & 0.12 &  0.06 \\ 
Eu II & 0.078 & 0.090 & $-$0.014 & 0.120 & 0.039 & 0.13 & 0.12 &  0.05\\
Gd II & 0.074 & 0.088 & $-$0.032 & 0.120 & 0.053 & 0.13 & 0.13 &  0.06 \\ 
Tb II & 0.078 & 0.091 & $-$0.016 & 0.121 & 0.029 & 0.12 & 0.11 &  0.04 \\
Dy II & 0.076 & 0.088 & $-$0.023 & 0.118 & 0.058 & 0.13 & 0.13 &  0.07 \\     
Er II & 0.078 & 0.080 & $-$0.088 & 0.142 & 0.040 & 0.15 & 0.15 &  0.09 \\ 
Tm II & 0.092 & 0.053 & $-$0.043 & 0.115 & 0.058 & 0.13 & 0.14 &  0.08 \\ 
Hf II & 0.042 & 0.086 & $-$0.028 & 0.123 & 0.170 & 0.21 & 0.20 &  0.17 \\ 
Os I  & 0.078 & 0.011 & $-$0.008 & 0.157 & 0.120 & 0.20 & 0.19 &  0.16 \\ 
Ir I  & 0.065 & 0.045 & $-$0.024 & 0.140 & 0.090 & 0.17 & 0.15 &  0.11 \\ 
Th II & 0.048 & 0.090 & $-$0.008 & 0.132 & 0.020 & 0.13 & 0.11 & \dots \\ 
 U II & 0.046 & 0.093 & $-$0.002 & 0.131 & 0.110 & 0.17 & \dots &  0.11 \\ 
\hline\hline
\end{tabular}
\end{center}
\end{table*}

\subsection{The lighter elements, 38$\leq$Z$\leq$48}

This group includes the classically observed elements Sr, Y, and Zr, but also
the less well-studied species Nb, Ru, Rh, Pd, and Ag, whose lines are weak and
lie in the near-UV part of the spectrum, hampering their detection in normal
metal-poor halo giants.  The only metal-poor star in which all 
these elements have been previously detected is \sned (Sneden et al.
\cite{sneden00}), thanks to its large enhancement of neutron-capture elements.
In \cs we detected even more lines of these elements (due to the slightly 
larger metallicity, reduced blending by CH, CN and NH molecules, and better spectrum quality); Figs. \ref{F-Nb-Ru-Rh}
and \ref{F-Pd-Ag} show
examples of the quality of the fits obtained. However, one element detection 
remains inconclusive -- even the strongest expected transition of Cd~I in
our wavelength domain (3261.05 \AA) is too severely blended to be useful for
abundance determinations.
 
In Solar System material, the four lighter elements are dominated by
products of the main s-process (with a possible contribution from the weak
s-process as well; see Prantzos et al. \cite{prantzos90}, Tsujimoto et al.
\cite{tsujimoto00}), while the elements Ru, Rh, Pd, and Ag may contain a large 
r-process fraction (54\% to 86\%). However, in a star as metal-poor as \cs,
it is expected that the s-process contribution should be negligible, both from
a theoretical point of view (the main s-process takes place in AGBs, where
the contribution depends on metallicity via the abundances of both seed
nuclei and neutron sources, e.g.  Prantzos et al. \cite{prantzos90}), and from
an observational point of view (Burris et al. \cite{burris00} and references
therein). Thus, all these elements should represent r-process material
produced in the early Galaxy.  

In Fig. \ref{F-r-sun} we compare the observed
abundance pattern in \cs to the Solar System r-process abundances, scaled to
the abundance of \cs (see section \ref{second-peak} for details of the scaling
procedure). These Solar System r-process abundances are obtained from a
de-composition of the Solar abundances (Anders \& Grevesse \cite{grevesse89})
into their s- and r-process fractions, by subtraction of computed main
s-process yields (AGB yields) from the total abundances to obtain the r-
process fraction. We show here two sources for this de-composition,
illustrating the uncertainties involved.  The dashed line in Fig.
\ref{F-r-sun} follows the compilation of Burris et al. (\cite{burris00}), which
uses yields from K\"appeler et al.  (\cite{kappeler89}) and Wisshak et al.
(\cite{wisshak96}), while the solid line uses yields of AGB models from
Arlandini et al.  (\cite{arlandini99}).

It is clear from inspection of Fig. \ref{F-r-sun} that, although one can
argue the case for general agreement in the region of the second r-process
peak, the abundances of \cs in the region 38$\leq$Z$\leq$48 are not all
compatible with the Solar System r-process pattern. This effect is best seen in
the middle and lower panels of Fig. \ref{F-r-sun}, where the abundance
difference $\log \epsilon_{*} - \log \epsilon_{rSS}$ between \cs and the Solar 
System (SS) $r$-process are displayed. When the Burris el al. (\cite{burris00})
de-composition is used, the difference appears as a stronger odd-even effect in
\cs, in addition to a lower mean abundance: $<\log \epsilon_{*} - \log
\epsilon_{rSS}>_{38\leq Z\leq 48}=-1.47$ (rms 0.33) for the lighter elements 
vs. $<\log \epsilon_{*} - \log \epsilon_{rSS}>_{56\leq Z\leq 69}=-1.25$ 
(rms 0.10) for the heavier group.  
If the Arlandini et al. (\cite{arlandini99}) de-composition is used,
the result is very similar for Ru, Rh, and Ag.  Note that in this
case, the Y abundance is no longer discrepant, while Nb is significantly
more abundant than the Solar r-value.  As a result, the mean offsets
between the lighter and second-peak elements over large intervals
in atomic number are in fact quite similar,  $<\log \epsilon_{*} - \log
\epsilon_{rSS}>_{38\leq Z\leq 48}=-1.38$ (rms 0.33) compared to $<\log
\epsilon_{*} - \log \epsilon_{rSS}>_{56\leq Z\leq 69}=-1.28$ (rms 0.08).

Independently of the decomposition used, the most
discrepant element is silver, for which the solar system r-process
scaled abundance exceeds the \cs observed value by 0.8 dex ! This low
abundance of Ag was also observed in \sned by Sneden et
al. (\cite{sneden00}), and the good agreement
between the \cs and \sned silver abundances can be seen 
in Fig. \ref{F-r-cs22}).
In contrast, the only other metal-poor stars (4 halo stars with
$-2.15\leq$[Fe/H]$\leq -1.3$ dex) in which it was observed so
far (Crawford et al. \cite{crawford98}) demonstrated a mild [Ag/Fe]
enhancement, in agreement with the mild enhancement of the second
r-process peak elements in these stars. This illustrates that not
everything is understood in the way the r-process builds up elements in
the wide atomic mass-range in which it is at work.

\subsection{The second-peak elements, 56$\leq$Z$\leq$72}\label{second-peak}

The group of elements between barium and hafnium is the best studied mass-range
of the neutron-capture elements, thanks to the relatively strong lines in the
visible region of elements such as Ba, Eu, and La. In \cs we were able to 
detect lines from {\it all} stable elements between Z=56 and Z=72. However,
three of them cannot be used for abundance determinations because of poorly
known atomic physics --  Ho and Lu have very strong hyperfine structure which
is not well quantified; and while three lines of Yb were detected, two
are severely blended (3289\AA and 3476\AA), and the strongest one
(3694\AA) yields a very large abundance ($\log\epsilon$(Yb)=0.18),
which we believe is due to the unaccounted hyperfine
structure, which acts to de-saturate this strong line (123 m\AA). 
We are thus left with accurate abundance determinations for 13
elements in the second peak of the r-process. 
Fig. \ref{F-HfEu} are example of the fit respectively of a Hf and
a Eu line. Not that the europium hyperfine structure used here is from
Kurucz (Kurucz \cite{kuruczcd}), although the oscillator strength was
taken from the more recent work of Lawler et
al. \cite{lawler01b}. The Eu isotope composition
adopted was 47.8$\%$ of $^{151}$Eu and 52.2$\%$ of 
$^{151}$Eu, as in the solar system, and in accordance to the new
measurement of Sneden et al. (\cite{snedenEu}) who measured this
isotopic ratio to be solar in the two metal-poor r-process rich stars
BD\,+17\,3248, \hd115 and \sned. The europium isotopic composition of \cs
will be investigated in a forthcoming paper, together with the rest of
our sample of extremely metal poor stars.

The abundances of the second peak elements are displayed in Fig.
\ref{F-r-sun}, and compared to the Solar System r-process, scaled by the mean
abundance difference with \cs: $<\log \epsilon_{*} - \log
\epsilon_{rSS}>_{56\leq Z\leq 69}$.  Both the Burris et al. (\cite{burris00})
and the Arlandini et al. (\cite{arlandini99}) de-compositions are very similar
in this mass range, and the mean underabundances computed for \cs are 
$<\log \epsilon_{*} - \log \epsilon_{rSS}>_{56\leq Z\leq 69}=-1.25$ (rms 0.10) 
and $-1.28$ (rms 0.08), respectively.
The remarkable agreement of the abundance ratios in halo stars with the Solar
r-process pattern in this atomic mass-range has been noted already in several
papers (Sneden et al. \cite{sneden96},\cite{sneden00}; Westin et al.
\cite{westin00}; Johnson \& Bolte \cite{johnson01}), and is also seen
in giants of the globular cluster M15 (Sneden et al. \cite{snedenM15}). In this
respect \cs resembles other metal-poor stars, both mildly r-process-enhanced
(\hd115 and others, see Johnson \& Bolte \cite{johnson01}), and the extreme
r-process-enriched star \sned.  In fact, Fig. \ref{F-r-cs22} shows that the
neutron-capture-element pattern in \cs (this paper) and \sned\, (Sneden et
al. \cite{sneden00}) are virtually indistinguishable (\sned\, abundances have
been scaled by the mean difference between the two stars $<\log \epsilon_{\cs}
- \log \epsilon_{\sned}>_{56\leq Z\leq 69}=+0.17$ (rms 0.10)).  Note  that
while the absolute r-process abundances are larger in \cs, given the
metallicity difference between the two stars ([Fe/H]$_{\cs}=-2.9$ and
[Fe/H]$_{\sned}=-3.2$), the total [r/Fe] ratio in \cs\, in the mass range
56$\leq$Z$\leq$69 is 0.13 dex lower than in \sned.
     
\subsection{The third-peak elements and the actinides, 
76$\leq$Z$\leq$92}\label{actinides}

The third r-process peak (near the magic number N=126) is sampled in
\cs by Os and Ir. The two heaviest species detected are the radioactive
actinides Th and U, the use of which as chronometers we discuss in 
section \ref{age}.

   \begin{figure}
   \centering
   {\includegraphics[angle=0,width=4.2cm]{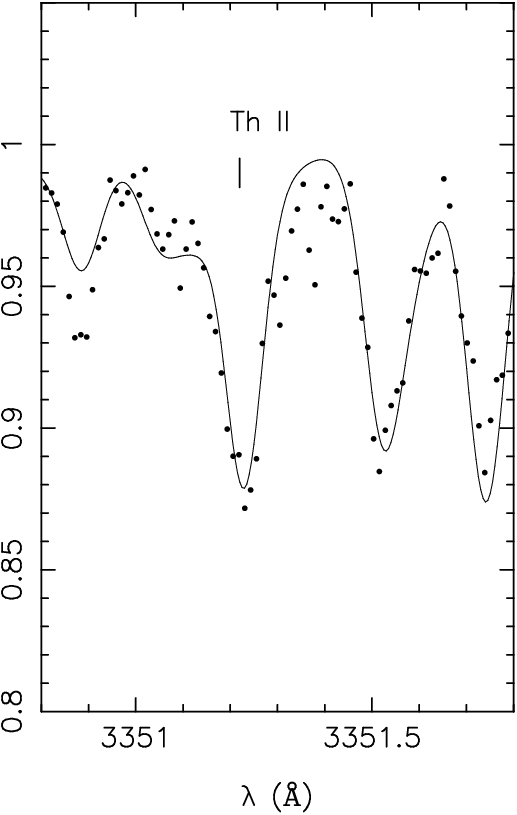}
   \includegraphics[angle=0,width=4.2cm]{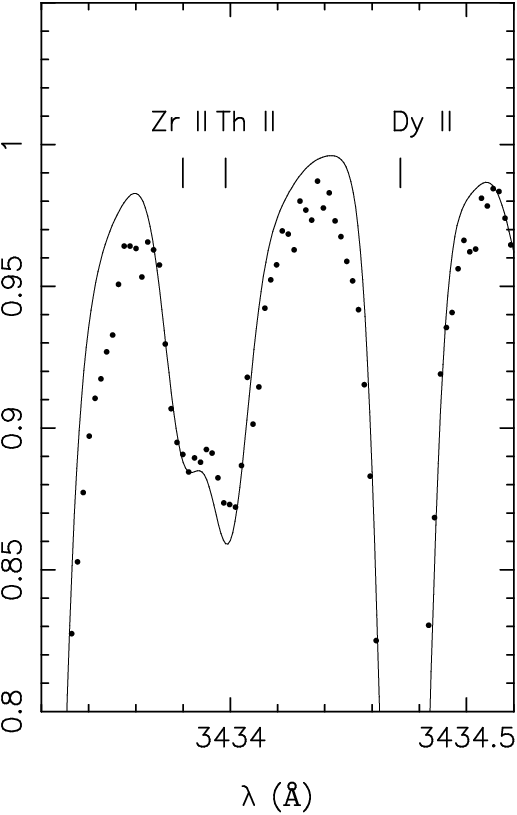}
   \includegraphics[angle=0,width=4.2cm]{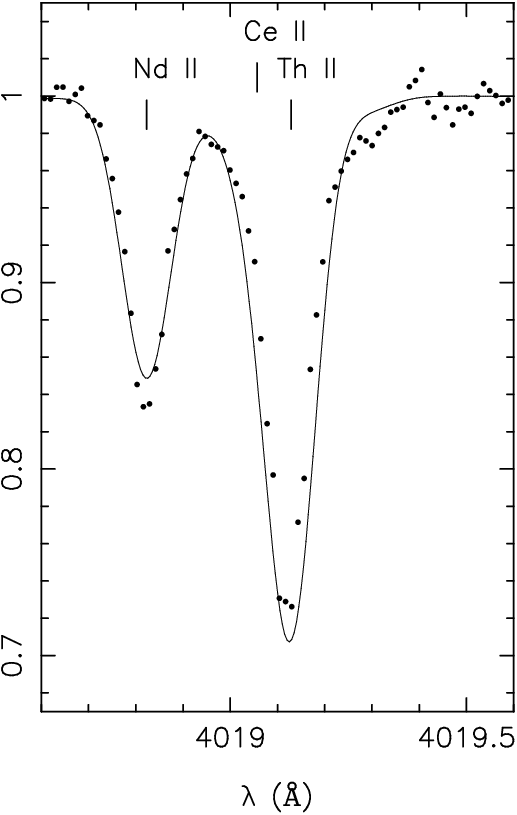}
   \includegraphics[angle=0,,width=4.2cm]{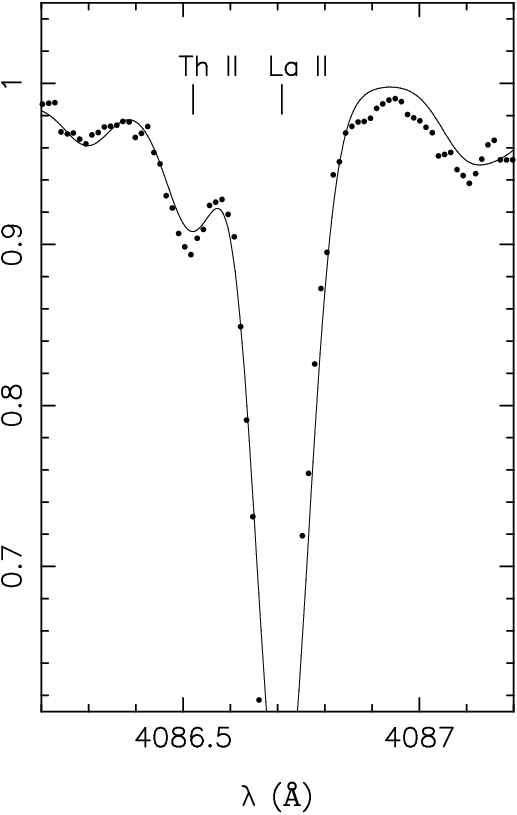}}
      \caption{The observed Th II 3351 \AA, 3434\AA, 4019\AA\, and 4086\AA\  
      lines in \cs. {\it Dots:} observations; {\it line:} synthetic spectrum
      computed for the mean Th abundance, $\log \epsilon$(Th)=-0.98.}
         \label{F-Th}
   \end{figure}

   \begin{figure*}
   \centering
   {\includegraphics[angle=0,height=7.5cm]{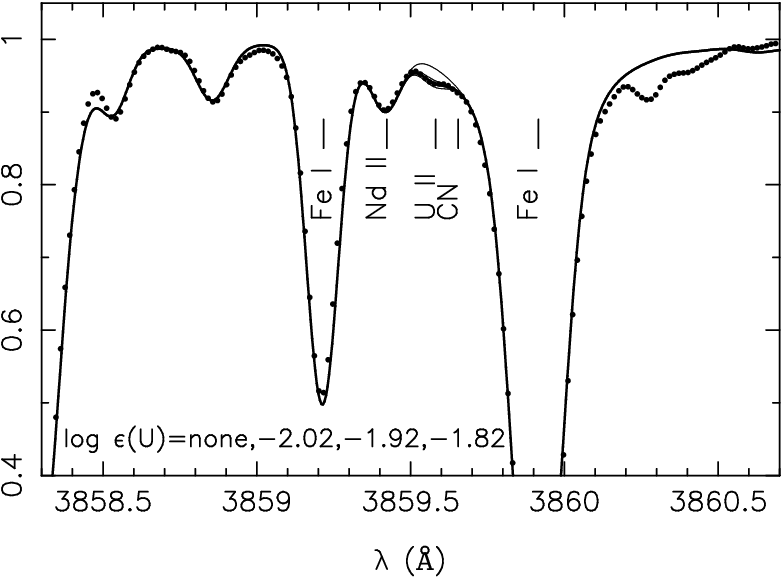}
    \includegraphics[angle=0,height=7.5cm]{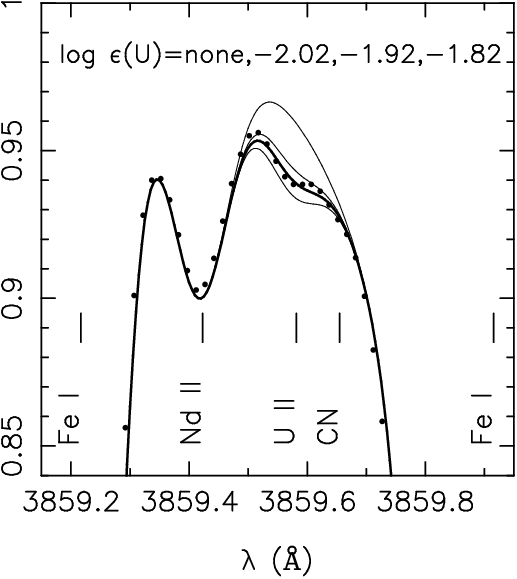}}
    \caption{The observed U~II 3859 \AA\ line in \cs. Symbols as in Fig.
      \ref{F-C}. The best fit is found for $\log \epsilon$(U)=$-$1.92 
      (thick line).}
         \label{F-U}
   \end{figure*}

\noindent{\it Osmium and Iridium}\\
Fig. \ref{F-Os-Ir} shows two of the three osmium detections, and one of the
two iridium detections. It was suggested in our preliminary results (Hill et
al. \cite{hill01}) that these two elements were overabundant with respect to
the Solar r-process by around +0.3 dex.  We now revise this statement slightly.
In particular, Ir falls back to the same abundance scale as the $56\leq
Z\leq 69$ elements. The reason for this revision is connected with the code
used to derive abundances. In our preliminary analysis (Hill et al.
\cite{hill01}) we were using the code of Spite (\cite{codespite}), 
while
in the present analysis we have switched to a more self-consistent approach,
the synthesis code by Plez et al. (\cite{codeplez}), which employs the same
algorithms to compute the model atmosphere and the synthetic spectrum. The
main difference between the two codes lies in the continuous opacity
computations and the source function assumptions (a diffusive term is added in
the latter). These codes provide identical results above 4000 \AA\ (less
than 0.02 dex difference), but the present, presumably more reliable  
approach, yields systematically lower
abundances in the bluest part of the spectrum (with a maximum effect of
$\sim$0.2 dex).  In the case of Ir, the two lines at 3512 \AA\ and 3800 \AA\
gave discrepant results in our preliminary analysis, but now agree, with $\log
\epsilon(Ir)=0.2$ dex (instead of the earlier 0.37 dex which came from
the 3512 \AA\ line alone).

Osmium, on the other hand, still seems to be overabundant with respect
to both the scaled Solar r-process fraction and to \sned. If Os in \cs\, 
was enhanced
to the same level as the second r-process peak elements, we would expect
$\log \epsilon$(Os)=0.15 dex, and $\log \epsilon$(Os)=0.12 dex if it was
enhanced similarly to \sned, whereas we observe $\log
\epsilon$(Os)=0.43 dex (rms 0.17, from 3 lines). 
We investigated the
possible source of differences between our analysis and that of
Sneden et al. (\cite{sneden96}, \cite{sneden00}), which could explain
this difference: (i) The three lines used
here are the same as those used by Sneden et al. (same wavelengths, same excitation potential and same oscillator strengths), 
(ii) the ionisation potential of Os I adopted
here is from a measurement by  Colarusso et al. (\cite{colarusso97})
of 8.44eV, compared to 8.35eV adopted by Sneden et al. (\cite{sneden96}),
which could account for at most $\sim$0.1 dex difference in the final
abundance but in the wrong direction, (iii) the difference in the codes
used for the abundance determination cannot play a role {\it only for
osmium}, leaving all the other abundance determinations unaffected. 
Hence, at this point, we cannot account for the abundance discrepancy
by any obvious differences in the analysis. We are therefore left with the
possibility that the Os content of \cs is indeed larger than predicted
by a scaled Solar r-process. On  the other hand, the large abundance
dispersion observed from the three lines (rms 0.17) is a hint that
there may be hidden problems in the determination of Os abundance from
these lines, so that any strong conclusion would be premature at this stage.

\noindent{\it Lead}\\
Recently, accurate abundances of lead in metal-poor CH stars (the Pb having 
likely been transferred from a now-extinct AGB companion) were reported
by Aoki et al. (\cite{aoki00}) and Van Eck et al.  (\cite{vaneck01}), using the
4057.8 \AA\ line.  In \cs, this line is not visible. Hence, we can only assign 
an upper limit of $\log \epsilon$(Pb)$<-0.2$ dex ($\log \epsilon$(Pb)$=
-0.4^{+0.2}_{-\infty}$), as shown in Fig. \ref{F-Pb}. However, even this upper
limit is of great interest, since it is already {\it below} the expected
abundance of the scaled Solar System r-process fraction (Fig. \ref{F-r-sun}).

Noting from Fig. \ref{F-r-cs22} that our derived abundance of Pb is drastically
lower than the detection reported by Sneden et al. (\cite{sneden00})
for \sned,  
we re-assessed also the abundance of Pb in \sned from spectra taken
with VLT+UVES during the commissioning of the instrument
(http://www.eso.org/science/uves\_comm/). The spectrum was acquired
with a resolution of R=55,000; the S/N of the co-added spectrum 
(total exposure time of 4.5h) is $\sim$140 per (0.013\AA) pixel at 4100\AA\ 
(i.e., S/N$\sim$330 per resolution element).
The model for \sned is an OSMARCS
model with $T_{eff}=$4700K, $\log g=$1.5, [Fe/H]=-3.0 and [$\alpha$/Fe]=+0.40
(following Sneden et al. \cite{sneden00}).  The syntheses were generated with
$\xi$=2.1 kms$_{-1}$, [Fe/H] = $-$3.2, 
and individual abundances from Sneden et al.
(\cite{sneden96}). C and N abundances were determined, through a fit of the
3850-3900 \AA\ region (CH and CN bandheads), to be  $\log
\epsilon$(C)=6.07 ([C/Fe]=+0.75) and $\log \epsilon$(N)=5.42 ([N/Fe]=+0.70).
Then the Pb region was synthesized (some gf values of nearby atomic lines were
adjusted to fit the observed spectrum).  The Pb line lies in the red wing of a 
CH line, which is nicely fitted with the abundances derived from the
3850--3900 \AA\ region. Spectra were computed for $\log
\epsilon$(Pb)=$-$0.5, 0.0, 0.5,  
and no Pb. From inspection of Fig. \ref{F-Pb} we derive an upper limit for 
the Pb abundance of
$\log \epsilon$(Pb)$< 0.0$ ($\log \epsilon$(Pb)$= -0.25^{+0.25}_{-\infty}$).
The Pb contents of \sned\, and \cs, therefore, do not seem to be very
different,  and the reality of the measurement by Sneden et
al. (\cite{sneden00}) appears open to question.

\noindent{\it Thorium and Uranium}\\
The oscillator strengths of the single U~II line and eight of the Th~II lines
that we have measured in \cs (Table \ref{T-Th}), as well as many others, have
been re-determined with superior accuracy by Nilsson et al. (\cite{nilsson01a},
\cite{nilsson01b}). The associated change in the U/Th abundance ratio is quite
significant, as the oscillator strengths of the Th~II lines decrease by 0.07
dex, on average, whereas the $\log gf$ of the U~II line increases from $-$0.20
to $-$0.067, i.e., by 0.13 dex.  Moreover, the uncertainties associated with
the oscillator strengths have been reduced drastically (to better than 0.08
dex for the individual Th lines, and to only 0.06 dex for the U~II line), thus
they are a negligible source of error compared to the uncertainties in the
actual fit of the data.  Fig. \ref{F-Th} displays the synthesis of a
selection of the observed thorium lines, all plotted with a thorium abundance
equal to the mean of the eight lines ($\log \epsilon$(Th)=$-0.98$). 
The observed
uncertainty in this case was estimated as $\sigma/\sqrt{N-1}$, where $\sigma$
is the dispersion around the mean and N, the number of lines, hence leading to
a $\log \epsilon(Th)=-0.98\pm 0.02$.  

Fig. \ref{F-U} shows the region of the
U~II 3859 \AA\ line and an enlargement of the uranium line itself, together
with synthetic spectra for four different uranium abundances.  The accuracy of 
the fit is estimated to be around 0.1 dex, obtained by testing the influence 
of several potential sources of error on the fitting procedure, including 
placement of the continuum and blending by neighboring lines (mainly Fe~I
3859.9 \AA). The uncertainty arising from the blending of
the Fe~I line is linked to uncertainties on the oscillator strength,
but also the broadening factor of the line. The Van der waals
broadening factor was taken from Barklem et al. \cite{barklem2}, and the
oscillator strength was increased by 0.1 dex with respect to the VALD2
recommended value to give
a best fit to the observed line Fe~I 3859.9~\AA. 
Attempts to vary the Barkelem damping constant by 5\% or more induced
changes in the U abundance of less than 10\%. The unidentified 
features on the red side of the line (at 3860.77 and 3860.9~\AA\
respectively) hamper the cosmetics of the fit but have no influence on the uranium line region. We would like to
point out that the 3860.77~\AA\ unidentified line also appears in the
spectrum of \sned, whereas
it does not in other giants of similar metallicity and temperature
which have no excess of neutron-capture elements. We
therefore tentatively attribute this absorption feature to an
neutron-capture element, and encourage atomic physicists to work on
the identification of this feature. We also note here that there are
numerous features in the whole UV part of the spectrum of
neutron-capture enriched giants, that are still in need of identification.
A more detailed analysis of the 5-line feature (Fe~I 3859.21, 
Nd~II 3859.4, U~II 3859.57, CN 3859.67 and Fe~I 3859.9~\AA) 
is forseen is a near future, involving 3D hydrodynamical models.
Adding the 0.06 dex uncertainty associated with the $\log gf$ of
the line results in $\log \epsilon(U)=-1.92\pm 0.11$. 
As seen from Table \ref{T-error}, the overall uncertainty of the U/Th ratio is 
totally dominated by that of the U line fitting procedure, while errors in the 
stellar parameters cancel out completely.

Finally, we note that, using the newly determined oscillator strength value for 
the uranium line, the upper limit deduced by Gustafsson \& Mitzuno-Wiedner
(\cite{gustafsson01}) for \sned\, becomes $\log \epsilon$(U)$\leq -2.54$.

\section {Is the r-process universal ?} \label{r universal}

   \begin{figure}
   \centering
   {\includegraphics[angle=-90,width=8.5cm]{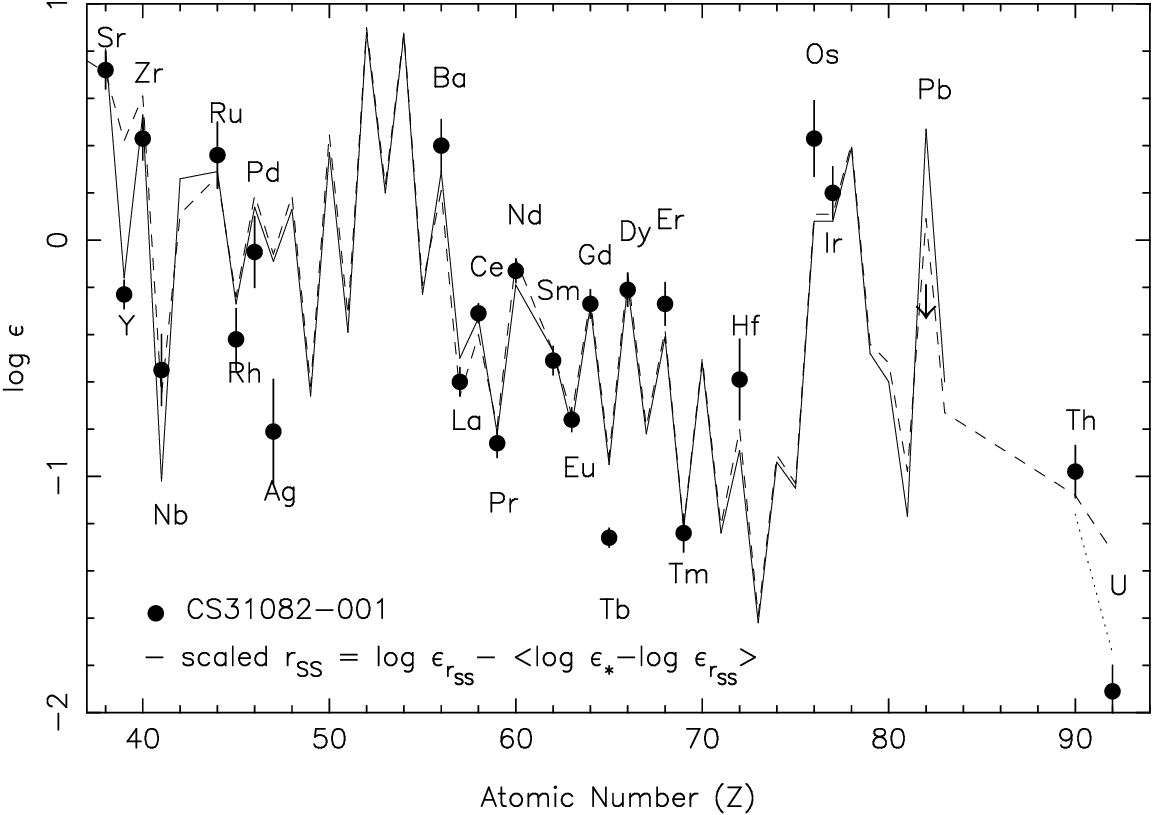}
    \includegraphics[angle=-90,width=8.5cm]{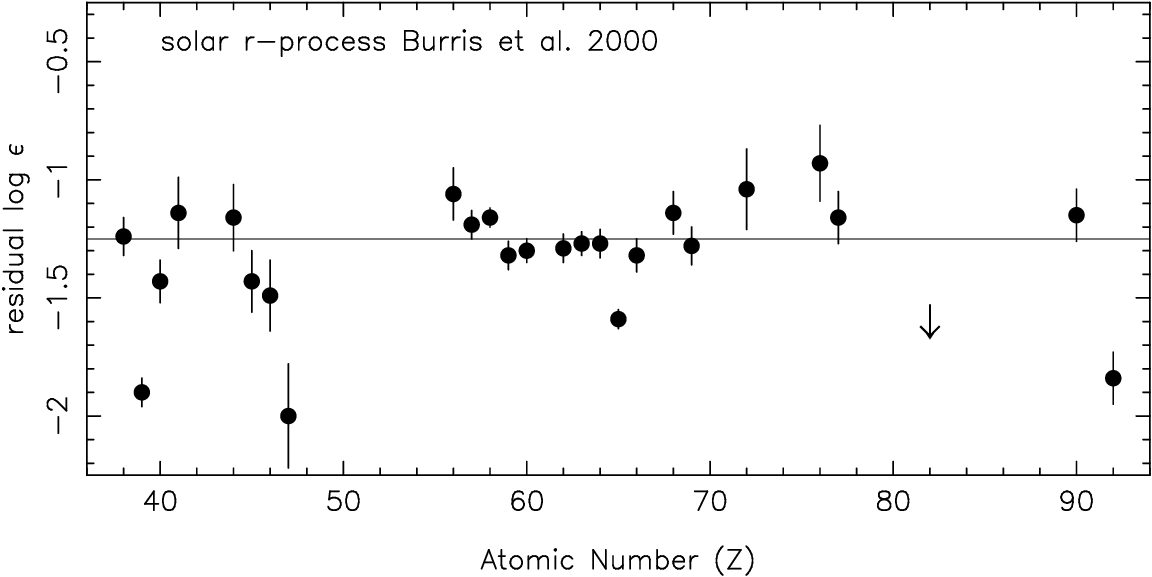}
    \includegraphics[angle=-90,width=8.5cm]{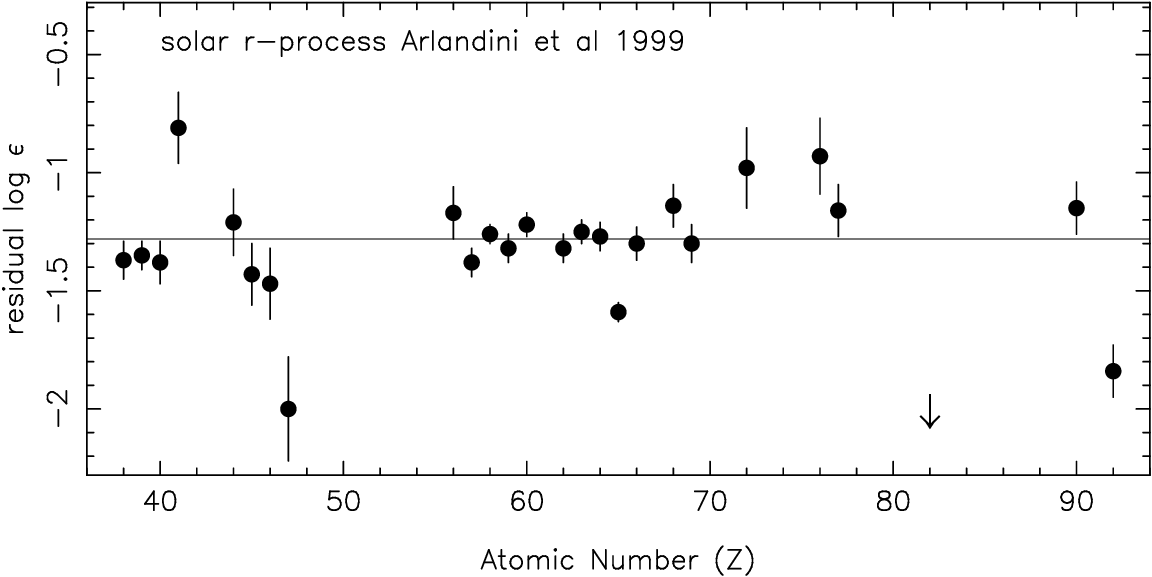}}
      \caption{{\it Top:} Neutron-capture-element abundances of \cs compared 
to the Solar System r-process scaled to match the 
56$\leq$Z$\leq$69 elemental abundances 
of \cs. Two sources are plotted for the Solar r-process: Burris et al. 
\cite{burris00} (dashed line) and Arlandini \cite{arlandini99} (full 
line). Note that the radioactive species (Th and U) Solar System abundances are corrected for
radioactive decay since the formation of the Solar System. The dotted line
show the abundances observed {\it today} for these two species (scaled
in the same manner).
{\it Middle:} residual abundance of \cs after the Solar System
      r-process (Burris et al. \cite{burris00}) has been subtracted. 
{\it Bottom:} residual abundance of \cs after the Solar System
      r-process (Arlandini et al. \cite{arlandini99}) has been subtracted.}
         \label{F-r-sun}
   \end{figure}

\begin{figure}
   \centering
   {\includegraphics[angle=-90,width=8.5cm]{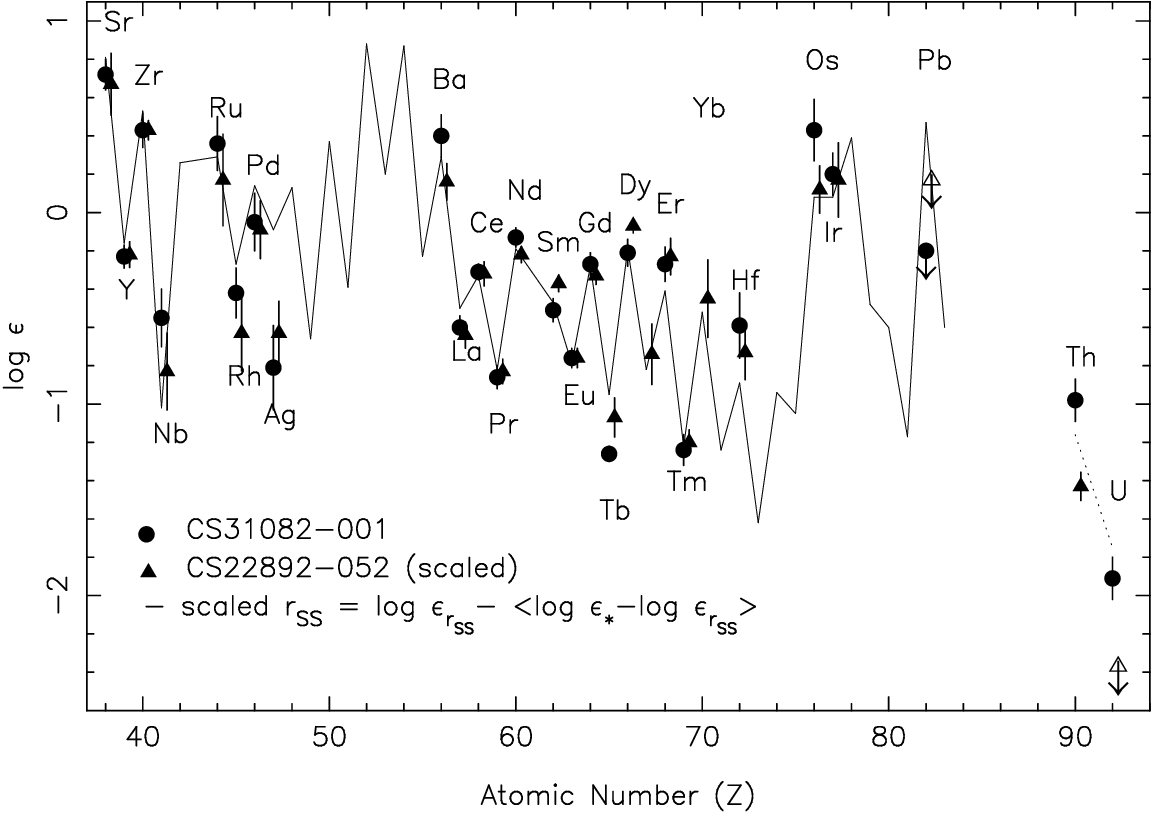}}
      \caption{Neutron-capture-element abundances of \cs compared to \sned
(Sneden et al. \cite{sneden00}. The
abscissa for \sned has been artificially shifted by +0.3 for readability, and
the abundances were scaled up by $<\log \epsilon_{\cs}
- \log \epsilon_{\sned}>_{56\leq Z\leq 69}=+0.17$dex. The two open symbols are
our own estimates for the Pb and U content of \sned (see section 
\ref{actinides}). 
The full line is the Solar r-process fraction from Arlandini et al. 
\cite{arlandini99}.}
         \label{F-r-cs22}
   \end{figure}

From the discussion in the previous section, and from Figs. \ref{F-r-sun} and
\ref{F-r-cs22}, it is clear that the neutron-capture elements in \cs follow a
standard pattern (i.e., they are indistinguishable from both the scaled 
Solar r-process pattern and the patterns observed in other
metal-poor halo stars such as \sned) for elements 56$\leq$Z$\leq$72. The small
dispersion around the mean of the quantities ($\log \epsilon_{\cs} - \log
\epsilon_{\sned}$) and ($\log \epsilon_{\cs} - \log \epsilon_{r_{SS}}$)
in the range 56$\leq$Z$\leq$69  (respectively 0.10 and 0.08 dex) reflect this
level of agreement.

In \cs, the third neutron-capture peak is so far only sampled by
abundance measurements of two elements (Os and Ir), and 
one upper limit (Pb).  
There may be marginal
evidence for departure from the Solar r-process (Os seems
overabundant), 
but it is premature to conclude firmly on this point 
(see section \ref{actinides}).
The third-peak abundance determinations clearly demand
confirmation from better measurements and laboratory data (Os), and from new 
detections (Pt, Pb, Bi), which can only be done from space, as the strongest 
lines of these elements are too far in the UV region to be reached from the 
ground.

On the other hand, the actinides, although only probed by the two radioactive
nuclides Th and U, do appear to be enhanced in \cs\, to a higher level than
observed for elements of the second r-process peak.  Given the very high ratios
of $\log$(Th/Eu)=$-$0.22 dex (where Eu is taken as a typical example of the
elements 56$\leq$Z$\leq$69) compared to other halo stars (for example, \sned\,
with $\log$(Th/Eu)=$-$0.66, and
\hd115 with $\log$(Th/Eu)=-0.60) it is difficult to conceive any reasonable 
scenario that would account for this by an age difference: \sned and \hd115 
would then be 20 and 18 Gyrs older than \cs, respectively (regardless of the 
adopted production ratio for Th/Eu), which seems unrealistic.  

We are thus left with the possibility that the actinides were enhanced 
{\it ab initio} by a larger factor than the elements of the second r-process
peak in the matter that gave birth to \cs.  This is the first time that such a
large departure ($\sim$0.4 dex) from the otherwise standard Solar r-process 
pattern has been observed in a halo star, and the implications are important. 
{\it If the actinides are not necessarily produced together with the lighter 
neutron-capture elements (56$\leq$Z$\leq$72), and their initial proportions 
are therefore not fixed, but instead vary from star to star, then any 
chronometer based on ratios of an actinide to any stable element from the 
second r-process peak is doomed to failure.} The Th/Eu ratio
in \cs\, is an extreme example of such a failure (see section \ref{age}). 

From the r-process modeling point of view, the de-coupling of the production of
actinides from the production of lighter r-process elements is in fact not 
unexpected.   Goriely \& Arnould (\cite{ga01}) find in their superposition of
CEVs (Canonical EVents) that reproduce the Solar r-process pattern, that {\it
the CEVs that are responsible for the synthesis of the actinides do not
contribute to the synthesis of nuclides lighter than Pb}. This point is considered in more detail by Schatz et al. (\cite{schatz02}).

\begin{figure}
   \centering
   {\includegraphics[angle=-90,width=8.5cm]{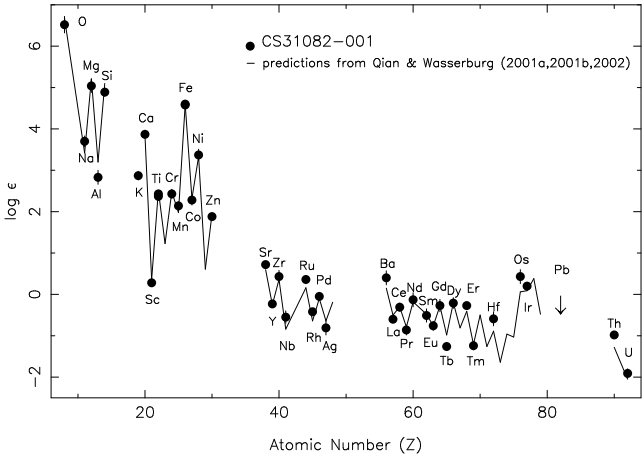}}
      \caption{Abundances of \cs compared to the predictions of Qian \&
      Wasserburg (\cite{QW01b},\cite{QW02}) three component
      phenomenological model.}
         \label{F-QW}
   \end{figure}

As a final remark, we compared abundances of all elements from Na
to U, to the predictions of the Qian \& Wasserburg
(\cite{QW01b},\cite{QW02}) phenomenological model which describes the
chemical enrichment in the early galaxy in terms of three components:
the prompt enrichment (P) is the contribution from extremely massive
stars and acts on an instantaneous timescale, and SN\,II are divided
in two classes, the high-frequeny SN\,II (H) and low-frequency SN\,II
(L) are responsible respectively for the second and third r-process
peak elements, and for some iron and light r-process elements. As in
Qian \& Wasserburg (\cite{QW01a}), the total number of H events that
contributed to the abundances of \cs can be computed as $n_{H}\simeq
52$ from the observed Eu abundance in this star (predictions for the
other r-process elements are made according to empirical yields
determined from the observed neutron-capture elements abundances 
in \sned and \hd115), while the P component
is responsible for all elements Na-Zn (where the yields are computed
in Qian \& Wasserburg (\cite{QW02}) directly from the observed
abundances of extremely metal poor stars, \hd115 in this case).
The overall excellent agreement between the observed abundances in \cs
and the predictions of Qian \& Wasserburg seen in Fig. \ref{F-QW} is 
therefore showing primarily, (i) that the abundances Na-Zn of \cs are
in very good agreement with {\it normal} extremely metal poor stars;
(ii) that the neutron-capture pattern up to Z=70 is also in very good
agreement with their H yields (ie very close to the abundances of
\sned) and (iii) that the Os, Th and U abundances are above the
predictions, as previously discussed in this section.

\section{The age of \cs} \label{age}

The very low metallicity of \cs, well below that of the most metal-poor
globular clusters, shows that it was formed in the earliest star formation
episodes, either in the Galaxy or in a substructure which later merged with the
Galaxy.  Comparison with the metallicity of damped Ly-$\alpha$ systems at high
redshift (Lu et al. \cite{lu96}, Vladilo et al. \cite{vladilo00}) shows that
the matter in \cs\, probably originated at an epoch earlier than $z=5$.
Assuming $\Omega _0=0.3$, $H_0$ in the plausible range 65--75 km/s/Mpc, and a
flat geometry (de Bernardis et al. \cite{bernardis}), the Big Bang occurred
about 0.5 Gyr before the epoch $z=10$, and 1 Gyr before $z=5$. Accordingly, the
age of the progenitor of \cs, as determined from the U/Th clock, provides a
lower limit to the age of the Universe which is {\it very} close to that age
itself.

Dating matter from the decay of a radioactive isotope is simple in
principle, provided that the ratio of the radioactive nuclide to a stable
reference element produced at the same time can be inferred. Let $(R/S)_0$
be the initial production ratio of a radioactive nuclide, $R$, to a stable one,
$S$, and (R/S)$_{now}$ be the value of that ratio observed today. The time for
$R/S$ to decay by a factor of 10 is then $\tau_{10} = \tau / \log (2)$,
where $\tau$ is the half-life of $R$. For $^{232}$Th and $^{238}$U ($\tau$
= 14.05 and 4.47 Gyr, respectively), this yields the following expressions
for the time $\Delta t$ (in Gyrs) elapsed since the production of these elements:

\begin{equation}
\Delta t = 46.67[\log(\mathrm{Th/S}_0)-\log(\mathrm{Th/S}_{now})]
\end{equation}
\begin{equation}\label{UTheq}
\Delta t =14.84[\log(\mathrm{U/S}_0)-\log(\mathrm{U/S}_{now})]
\end{equation}

These relations show that, if the right-hand sides can be evaluated to a
realistic accuracy of 0.1 dex, the corresponding error on $\Delta t$ becomes
4.7 Gyr for Th and 1.5 Gyr for U, demonstrating immediately the huge 
advantage of U over the previously used Th in cosmochronometry. However,
finding a stable $r$-process element that permits a spectroscopic
determination of (R/S)$_{now}$ {\it and} a theoretical prediction of $(R/S)_0$,
with a {\it combined overall} error of 0.1 dex is a significant, unsolved
problem.  

Hence, we look to an obvious alternative.  Substituting Th for the
stable element $S$ in eq. (\ref{UTheq}) above, we obtain:

\begin{equation}\label{UT}
   \Delta t =21.76[\log(\mathrm{U/Th}_0)-\log(\mathrm{U/Th}_{now})]
\end{equation}

Thus, for a given uncertainty in the decay of the U/Th ratio, the error in
$\Delta t$ is 50\% larger than for U alone, but still a factor of two 
better than
for Th alone. Adopting the slightly radioactive, but structurally very similar
Th, as the reference element for U leads to great gains in accuracy of both
terms on the right-hand side of eq. (\ref{UT}).  First, the ionization and
excitation potentials of the atomic levels giving rise to the observed spectral
lines are similar, so that errors in the model atmospheres and assumed
parameters largely cancel in the ratio U/Th (see Table \ref{T-error}).  Second,
the initial production ratio of the neighboring nuclides $^{238}$U and
$^{232}$Th should be far less sensitive to variations in the poorly-known
characteristics of the neutron exposure(s) occurring in the explosion of
progenitor than the ratios of nuclides more widely separated in mass, such as
$^{232}$Th and $^{151-153}$Eu (Goriely \& Clerbaux \cite{goriely99}).

Much progress has already been made on both fronts since the publication of our
discovery paper (Cayrel et al. \cite{cayrel01}). First, the oscillator
strengths of the single U~II line and eight of the Th~II lines that we have
measured in \cs have been re-determined (section \ref{actinides}). The
change in the U/Th abundance ratio is quite significant, revising the $\log
$(U/Th) from the previous value of $-0.74\pm 0.15$ to $-0.94\pm 0.11$.  Using
the same initial production ratio as in Cayrel et al. (\cite{cayrel01}), this
leads to an age of almost 17 Gyr, 4.3 Gyr greater than that originally
published.  By contrast, use of the conventional Th/Eu chronometer (Cowan et
al. \cite{cowan99}) leads instead to a slightly negative(!), or at most a 
T-Tauri like age for \cs.

Fortunately, there has also been progress regarding predictions of the initial
production ratio for these elements. In a recent paper, Goriely \& Arnould
(\cite{ga01}) review in great detail the production of the actinides in the
light of the Solar System data. Although they conclude that no current solution
explains the Solar System data exactly, stretching the lower and upper limits
of the production ranges by just 0.1 dex makes 10 of the 32 cases they 
consider acceptable. The corresponding production ratios range from 0.48 to
0.54, with a mean of $0.50 \pm 0.02$, close to the value of 0.556
(from Cowan et al. \cite{cowan99}) cited by Cayrel et al. (\cite{cayrel01}).
Combined with our newly measured U/Th ratio, this leads to an age of $14.0\pm
2.4$ Gyr for the Th and U in \cs (where the error refers only to the
uncertainty on the observed U/Th ratio). This is a quite reasonable value,
inspiring some hope that the predicted production ratio of U/Th is fairly
robust.

At this point, it is impossible to assign a reliable error estimate to this
age, given the lack of observational constraints on the production ratio of
U/Th from other species that might have experienced the same neutron exposure.
Verification of the predicted abundances of the decay products Pb and Bi, the
direct descendants of U and Th, is therefore of particular importance. No
useful lines of these elements are measurable, even in our high-quality
VLT/UVES spectra of \cs , and we suspect that they may remain unmeasurable,
regardless of improvements in future ground-based facilities.  Fortunately,
time on the Hubble Space Telescope has been assigned to obtain a
high-resolution UV spectrum of \cs, where the strong resonance lines of these
species might indeed be detected.

Let us note also that using the upper limit $\log \epsilon$(U)$\leq -$2.54 of Gustafsson \& Mitzuno-Wiedner 
(\cite{gustafsson01}, modified to account for the new $gf$ of the U line) 
for uranium in \sned, and the thorium abundance 
$\log \epsilon$(Th)=$-$1.60 of Sneden et al. (\cite{sneden00}), 
the ratio U/Th$\leq -$0.94 in \sned is fully compatible with the one of \cs.
Hence, despite a difference in the overall thorium content of the two stars,
 their ages derived from the U/Th ratio is fully consistent.

\section{Conclusions and future work}\label{conclusion}

The apparent brightness, very low [Fe/H], and enormous $r$-process enhancement,
in combination with the lack of molecular blanketing, makes \cs a uniquely
favorable object for study of the {\it r}-process nucleosynthesis 
history in the early
Galaxy. The most conspicuous result of these characteristics is the
breakthrough in the measurement of the abundance of uranium in this very old
star, facilitated by our excellent VLT/UVES spectra, which enable the
determination of accurate abundances for 41 (plus 2 upper limits -N
and Pb- and 3 detections which could not be translated 
into accurate determinations
-Ho, Lu and Yb- due to a lack of input atomic physics) other elements as
well. But the importance of \cs extends further, as it has provided the first
solid evidence that variations in progenitor mass, explosion energy, distance
to dense interstellar clouds, and/or other intrinsic and environmental factors
preceding the formation of the extreme halo stars, may produce significantly
different $r$-process abundance patterns from star to star in the actinide
region (Z$\geq$90).  

A striking consequence of these variations is the {\it complete failure 
of the conventional Th/Eu chronometer} in \cs, assuming an
initial production ratio for the pair as in CS 22892-052, or as in the {\it
r}-process elements of the Solar System. No such problem is seen for the pair
U/Th, which leads to an age of $14.0\pm 2.4$ Gyr (not including systematic
errors in the initial production ratio). This suggests that the closely similar
masses and nuclear structures of $^{238}$U and $^{232}$Th lead to a more stable
production ratio between these two nuclides than, e.g., that of $^{232}$Th and
$^{151-153}$Eu. However, further studies are needed of the robustness of the U/Th
ratio produced by exposing iron nuclei to neutron exposures of various
strengths, as well as of the abundances of the daughter elements of their
decay, Pb and Bi.

Observations in \cs itself of the Pb and Bi created in the same neutron
exposure event as U and Th, and by their subsequent decay, will be
particularly crucial as constraints on the predicted production ratios of
all $r$-process elements.  Of equal importance is the continuation of searches
for new r-process-enhanced metal-poor stars, so that better measures of the
star-to-star variation in the observed patterns of the r-process elements, in
particular those of the third-peak and the actinides, may be obtained.

\begin{acknowledgements}
This research has made use of the Simbad database,
operated at CDS, Strasbourg, France. We are grateful to Pr. Johansson and his 
group for their timely response to our request for accurate lifetime 
measurements of uranium and thorium electronic levels.   
BN and JA thank the Carlsberg Foundation and the Swedish and Danish Natural
Science Research Councils for financial support for this work.  T.C.B
acknowledges partial support from grants AST 00-98549 and AST 00-98508 from
the U.S. National Science Foundation.
\end{acknowledgements}

\newpage

\appendix
\section{Line list and atomic data}

We list in this table all the lines of neutron-capture elements that
were used to derive abundances. The wavelengths, excitation potentials,
and oscillator strengths are listed, together with references for the
oscillator strengths. Also listed are the equivalent widths of the
lines in \cs, and the derived abundances. The word ``syn'' in column 5
denotes that spectral synthesis techniques were used rather than the
equivalent width to derive the abundance. 

\begin{table}[h]
\caption{Linelist, equivalent width and abundances for the
neutron-capture elements in \cs.}
\begin{tabular}{ccrccr}
\hline\hline
$\lambda$ & Exc. Pot. & $\log gf$ & Ref. & W & $\log \epsilon$ \\
  (\AA)   & (eV)      &           &      & (m\AA) &              \\ 
\hline
Sr II \\
\hline
4077.709 & 0.00 &    0.170 & 1 &  syn  &    0.70\\
4161.792 & 2.94 & $-$0.600 & 1 &  syn  &    0.75\\
4215.519 & 0.00 & $-$0.170 & 1 &  syn  &    0.70\\
\hline
Y II\\
\hline
3774.331 & 0.13 &    0.210 & 2 &  86.0 & $-$0.29\\
3788.694 & 0.10 & $-$0.070 & 2 &  83.0 & $-$0.14\\
3818.341 & 0.13 & $-$0.980 & 2 &  45.0 & $-$0.15\\
3950.352 & 0.10 & $-$0.490 & 2 &  69.4 & $-$0.16\\
4398.013 & 0.13 & $-$1.000 & 2 &  45.0 & $-$0.23\\
4883.684 & 1.08 &    0.070 & 2 &  42.4 & $-$0.27\\
5087.416 & 1.08 & $-$0.170 & 2 &  28.6 & $-$0.32\\
5123.211 & 0.99 & $-$0.830 & 2 &  10.6 & $-$0.30\\
5200.406 & 0.99 & $-$0.570 & 2 &  19.8 & $-$0.25\\
5205.724 & 1.03 & $-$0.340 & 2 &  26.0 & $-$0.28\\
\hline
Zr II\\
\hline
3836.762 & 0.56 & $-$0.060 & 3 &  61.0 &    0.24\\
4161.213 & 0.71 & $-$0.720 & 3 &  40.0 &    0.57\\
4208.977 & 0.71 & $-$0.460 & 3 &  45.0 &    0.40\\
4317.299 & 0.71 & $-$1.380 & 3 &  13.0 &    0.54\\
4496.980 & 0.71 & $-$0.590 & 3 &  35.0 &    0.30\\
\hline
Nb II\\
\hline
3215.591 & 0.44 & $-$0.190 & 4$^{a}$ &  syn  & $-$0.55\\
\hline
Ru I\\
\hline
3436.736 & 0.15 &    0.015 & 5$^{a}$ &  syn  &    0.40\\
3498.942 & 0.00 &    0.310 & 5$^{a}$ &  syn  &    0.28\\
3728.025 & 0.00 &    0.270 & 5$^{a}$ &  syn  &    0.20\\
3798.898 & 0.15 & $-$0.040 & 5$^{a}$ &  syn  &    0.40\\
3799.349 & 0.00 &    0.020 & 5$^{a}$ &  syn  &    0.45\\
\hline
Rh I\\
\hline
3396.819 & 0.00 &    0.050 & 4$^{a}$ &  syn  & $-$0.40\\
3434.885 & 0.00 &    0.450 & 4$^{a}$ &  syn  & $-$0.45\\
3692.358 & 0.00 &    0.174 & 4$^{a}$ &  syn  & $-$0.40\\
\hline
Pd I\\
\hline
3242.700 & 0.81 & $-$0.070 & 4$^{a}$ &  syn  &    0.00\\
3404.579 & 0.81 &    0.320 & 4$^{a}$&  syn  & $-$0.18\\
3634.690 & 0.81 &    0.090 & 4$^{a}$&  syn  &    0.00\\
\hline
Ag I\\
\hline
3280.679 & 0.00 & $-$0.050 & 4$^{a}$&  syn  & $-$0.95\\
3382.889 & 0.00 & $-$0.377 & 4$^{a}$&  syn  & $-$0.70\\
\end{tabular}
\end{table}

\begin{table}
\begin{tabular}{ccrccr}
\hline\hline
$\lambda$ & Exc. Pot. & $\log gf$ & Ref. & W & $\log \epsilon$ \\
  (\AA)   & (eV)      &           &      & (m\AA) &              \\ 
\hline
Ba II\\
\hline
3891.776 & 2.51 &    0.280 & 6 & syn  &    0.15\\
4130.645 & 2.72 &    0.560 & 6 & syn  &    0.20\\
4554.029 & 0.00 &    0.170 & 6 & syn  &    0.40\\
4934.076 & 0.00 & $-$0.150 & 6 & syn  &    0.60\\
5853.668 & 0.60 & $-$1.010 & 6 & syn  &    0.50\\
6141.713 & 0.70 & $-$0.070 & 6 & syn  &    0.40\\
\hline
La II\\
\hline
3849.006 & 0.00 & $-$0.450 & 7$^{a}$ & syn  & $-$0.60\\
4086.709 & 0.00 & $-$0.070 & 7$^{b}$ & syn  & $-$0.55\\
4123.218 & 0.32 &    0.130 & 7$^{b}$ & syn  & $-$0.65\\
5122.988 & 0.32 & $-$0.850 & 7$^{b}$ & syn  & $-$0.58\\
6320.376 & 0.17 & $-$1.520 & 7$^{b}$ & 7.0  & $-$0.61\\
\hline
Ce II\\
\hline
4073.474 & 0.48 &    0.320 & 8 &  26.3 & $-$0.45\\
4083.222 & 0.70 &    0.240 & 8 &  21.0 & $-$0.24\\
4120.827 & 0.32 & $-$0.240 & 8 &  16.0 & $-$0.37\\
4127.364 & 0.68 &    0.240 & 8 &  21.0 & $-$0.27\\
4222.597 & 0.12 & $-$0.180 & 1 &  33.0 & $-$0.25\\
4418.780 & 0.86 &    0.310 & 8 &  14.0 & $-$0.38\\
4486.909 & 0.30 & $-$0.360 & 1 &  22.0 & $-$0.14\\
4562.359 & 0.48 &    0.330 & 1 &  31.0 & $-$0.41\\
4628.161 & 0.52 &    0.260 & 1 &  26.6 & $-$0.40\\
\hline
Pr II\\
\hline
3964.262 & 0.22 & $-$0.400 & 8  &  syn  & $-$0.80\\
3964.812 & 0.05 &    0.090 & 9  &  28.0 & $-$0.99\\
3965.253 & 0.20 & $-$0.130 & 9  &  18.0 & $-$0.84\\
4062.805 & 0.42 &    0.330 & 10 &  37.0 & $-$0.70\\
5220.108 & 0.80 &    0.170 & 9  &   7.3 & $-$1.01\\
5259.728 & 0.63 & $-$0.070 & 9  &   8.1 & $-$0.93\\
\hline
Nd II\\
\hline
3973.260 & 0.63 &    0.430 & 11 &  syn  & $-$0.17\\
4018.823 & 0.06 & $-$0.880 & 12 &  19.0 & $-$0.20\\
4021.327 & 0.32 & $-$0.170 & 13 &  35.0 & $-$0.22\\
4061.080 & 0.47 &    0.300 & 11 &  62.0 &    0.07\\
4069.265 & 0.06 & $-$0.400 & 13 &  34.4 & $-$0.32\\
4109.448 & 0.32 &    0.180 & 11 &  syn  &   0.15\\
4232.374 & 0.06 & $-$0.350 & 13 &  34.5 & $-$0.39\\
4446.384 & 0.20 & $-$0.630 & 12 &  35.0 &    0.04\\
4462.979 & 0.56 & $-$0.070 & 11 &  38.0 & $-$0.03\\
5130.586 & 1.30 &    0.100 & 11 &  13.3 & $-$0.00\\
5212.361 & 0.20 & $-$0.700 & 13 &  14.6 & $-$0.48\\
5234.194 & 0.55 & $-$0.460 & 12 &  16.0 & $-$0.25\\
5249.576 & 0.98 &    0.080 & 14 &  20.0 & $-$0.16\\
5293.163 & 0.82 & $-$0.200 & 14 &  21.3 & $-$0.04\\
5311.453 & 0.99 & $-$0.560 & 14 &   5.4 & $-$0.16\\
5319.815 & 0.55 & $-$0.350 & 14 &  27.7 & $-$0.06\\
5361.467 & 0.68 & $-$0.400 & 13 &  12.6 & $-$0.29\\
5442.264 & 0.68 & $-$0.900 & 13 &   4.2 & $-$0.31\\
\hline
Sm II\\
\hline
3793.978 & 0.10 & $-$0.500 & 15 &  19.0 & $-$0.71\\
3896.972 & 0.04 & $-$0.580 & 15 &  21.0 & $-$0.66\\
4023.222 & 0.04 & $-$0.830 & 15 &  16.0 & $-$0.57\\
4068.324 & 0.43 & $-$0.710 & 15 &   7.0 & $-$0.65\\
4318.927 & 0.28 & $-$0.270 & 15 &  33.0 & $-$0.45\\
4499.475 & 0.25 & $-$1.010 & 15 &  13.0 & $-$0.30\\
4519.630 & 0.54 & $-$0.430 & 15 &  18.0 & $-$0.37\\
4537.941 & 0.48 & $-$0.230 & 15 &  16.0 & $-$0.71\\
4577.688 & 0.25 & $-$0.770 & 15 &  18.0 & $-$0.38\\
\end{tabular}
\end{table}

\begin{table}
\begin{tabular}{ccrccr}
\hline\hline
$\lambda$ & Exc. Pot. & $\log gf$ & Ref. & W & $\log \epsilon$ \\
  (\AA)   & (eV)      &           &      & (m\AA) &              \\ 
\hline
Eu II\\
\hline
3724.931 & 0.00 & $-$0.090 & 16$^{b}$ & syn  & $-$0.59\\
3930.499 & 0.21 &    0.270 & 16$^{b}$ & syn  & $-$0.78\\
3971.972 & 0.21 &    0.270 & 16$^{b}$ & syn  & $-$0.84\\
4129.725 & 0.00 &    0.220 & 16$^{b}$ & syn  & $-$0.77\\
4205.042 & 0.00 &    0.210 & 16$^{b}$ & syn  & $-$0.66\\
4435.578 & 0.21 & $-$0.110 & 16$^{b}$ & syn  & $-$0.76\\
4522.581 & 0.21 & $-$0.670 & 16$^{b}$ & syn  & $-$0.91\\
6437.640 & 1.32 & $-$0.320 & 16$^{b}$ & syn  & $-$0.88\\
6645.064 & 1.38 &    0.120 & 16$^{b}$ &  3.0 & $-$0.72\\
\hline
Gd II\\
\hline
3768.396 & 0.08 &    0.250 & 17 &  61.0 & $-$0.33\\
3796.384 & 0.03 &    0.030 & 18 &  63.0 & $-$0.13\\
3836.915 & 0.49 & $-$0.320 &  4$^{a}$ &  19.0 & $-$0.27\\
3844.578 & 0.14 & $-$0.510 & 17 &  30.0 & $-$0.22\\
3916.509 & 0.60 &    0.060 & 17 &  22.0 & $-$0.45\\
4037.893 & 0.73 &    0.070 & 18 &  15.0 & $-$0.53\\
4085.558 & 0.56 & $-$0.230 & 18 &  17.0 & $-$0.37\\
4130.366 & 0.73 &    0.140 & 18$^{a}$ &  25.0 & $-$0.32\\
4191.075 & 0.43 & $-$0.680 & 17 &  18.0 & $-$0.06\\
\hline
Tb II\\
\hline
3658.886 & 0.13 & $-$0.010 & 19$^{b}$ &  12.5 & $-$1.27\\
3702.853 & 0.13 &    0.440 & 19$^{b}$ &  33.0 & $-$1.19\\
3848.734 & 0.00 &    0.280 & 19$^{b}$ &  38.0 & $-$1.34\\
3874.168 & 0.00 &    0.270 & 19$^{b}$ &  29.0 & $-$1.19\\
3899.188 & 0.37 &    0.330 & 19$^{b}$ &  17.0 & $-$1.23\\
4002.566 & 0.64 &    0.100 & 19$^{b}$ &   4.0 & $-$1.38\\
4005.467 & 0.13 & $-$0.020 & 19$^{b}$ &  17.3 & $-$1.23\\
\hline
Dy II\\
\hline
3869.864 & 0.00 & $-$0.940 & 20 &  29.3 & $-$0.21\\
3996.689 & 0.59 & $-$0.190 & 20 &  32.6 & $-$0.20\\
4011.285 & 0.93 & $-$0.630 & 20 &   5.5 & $-$0.32\\
4103.306 & 0.10 & $-$0.370 & 20 &  61.2 & $-$0.01\\
4468.138 & 0.10 & $-$1.500 & 20 &   8.5 & $-$0.27\\
5169.688 & 0.10 & $-$1.660 & 20 &   5.5 & $-$0.38\\
\hline
Er II\\
\hline
3692.649 & 0.05 &    0.138 & 21 &  syn  & $-$0.25\\
3786.836 & 0.00 & $-$0.640 & 21 &  46.0 & $-$0.26\\
3830.482 & 0.00 & $-$0.360 & 21 &  58.5 & $-$0.26\\
3896.234 & 0.05 & $-$0.240 & 22 &  64.0 & $-$0.19\\
3938.626 & 0.00 & $-$0.520 &  8 &  38.0 & $-$0.40\\
\hline
Tm II\\
\hline
3700.256 & 0.03 & $-$0.290 & 8 &  25.0 & $-$1.28\\
3761.333 & 0.00 & $-$0.250 & 8 &  syn  & $-$1.10\\
3795.760 & 0.03 & $-$0.170 & 8 &  28.0 & $-$1.34\\
3848.020 & 0.00 & $-$0.520 & 8 &  syn  & $-$1.25\\
\hline
Hf II\\
\hline
3399.793 & 0.00 & $-$0.490 & 23 &  syn  & $-$0.50\\
3719.276 & 0.61 & $-$0.870 &  8 &  syn  & $-$0.70\\
\hline
Os I\\
\hline
4135.775 & 0.52 & $-$1.260 & 8 &   9.0+syn &    0.52\\
4260.848 & 0.00 & $-$1.440 & 8 &  12.0+syn &    0.21\\
4420.468 & 0.00 & $-$1.530 & 24 &  syn &    0.50\\
\hline
Ir I\\
\hline
3513.648 & 0.00 & $-$1.260 & 4$^{a}$ &  34.0+syn &    0.20\\
3800.120 & 0.00 & $-$1.450 & 4$^{a}$ &  syn  &    0.20\\
\hline
Pb I\\
\hline
4057.807 & 1.32 & $-$0.170 & 25$^{a}$ &  syn  &   $<$-0.2\\
\end{tabular}
\end{table}

\begin{table}
\begin{tabular}{ccrccr}
\hline\hline
$\lambda$ & Exc. Pot. & $\log gf$ & Ref. & W & $\log \epsilon$ \\
  (\AA)   & (eV)      &           &      & (m\AA) &              \\ 
\hline
Th II\\
\hline
3351.229 & 0.19 & $-$0.600 & 26$^{a}$ &   7.0+syn & $-$0.95\\
3433.999 & 0.23 & $-$0.537 & 26$^{a}$ &  10.0+syn & $-$1.05\\
3435.977 & 0.00 & $-$0.670 & 26$^{a}$ &  14.0+syn & $-$1.05\\
3469.921 & 0.51 & $-$0.129 & 26$^{a}$ &  10.0+syn & $-$1.00\\
3675.567 & 0.19 & $-$0.840 & 26$^{a}$ &   5.0+syn & $-$0.92\\
4019.129 & 0.00 & $-$0.228 & 26$^{a}$ &  31.0+syn & $-$1.03\\
4086.521 & 0.00 & $-$0.929 & 26$^{b}$ &  10.0+syn & $-$0.95\\
4094.747 & 0.00 & $-$0.885 & 26$^{a}$ &   7.0+syn & $-$0.95\\
\hline
U II\\
\hline
3859.571 & 0.036 & $-$0.067 & 27$^{a}$ & syn & $-$1.92 \\
\hline\hline
\end{tabular}

$^{a}$: not in Sneden et al. (\cite{sneden96}).

$^{b}$: $\log gf$ different than in Sneden et al. (\cite{sneden96}).

1: Gratton \& Sneden 1994;
2: Hannaford et al. 1982;
3: Bi\'emont et al. 1981;
4: VALD: Bell heavy; 
5: Wickliffe et al 1994;
6: Gallagher 1967;
7: Lawler et al. 2001a;
8: Sneden et al. 1996 (From Kurucz compilation);
9: Goly et al. 1991;
10: Goly et al. 1991, Lage \& Whaling 1976;
11: Maier \& Whaling 1977;
12: Ward et al. 1984, 1985, modified (Sneden et al. 1996);
13: Corliss \& Bozman 1962, modified (Sneden et al. 1996);
14: Maier \& Whaling 1977, Ward et al. 1984, 1985;
15: Biemont et al. 1989;
16: Lawler et al. 2001b;
17: Corliss \& Bozman 1962;
18: Bergstr\"om et al 1988;
18: Bergstrom et al 1988;
19: Lawler et al. 2001c;
20: Kusz 1992, Bi\'emont \& Lowe 1993;
21: Musiol \& Labuz 1993;
22: Bi\'emont \& Youssef 1984;
23: Andersen et al. 1975;
24: Kwiatkowski et al. 1984;
25: Reader \& Sansonetti 1986;
26: Nilsson et al. 2001a;
27: Nilsson et al. 2001b.

\end{table}
\end{document}